\documentclass[reprint,superscriptaddress,amsmath,amssymb,aps,floatfix,prb]{revtex4-1}
\usepackage[dvips]{graphics}
\usepackage{vmargin}
\usepackage{subfigure}
\usepackage{graphics}
\usepackage{epsfig}
\usepackage{dcolumn}
\usepackage{color}
\usepackage{bm}

\setpapersize{USletter}
\setmarginsrb{1in}{1.5in}{1in}{0.5in}{0in}{0.2in}{0in}{0in}

\begin{document}

\title{Electric control of
a $\{Fe_4\}$ single-molecule magnet in a single-electron
transistor}

\author{J. F. Nossa}
\affiliation{Department of Physics and Electrical Engineering, Linnaeus University, SE-39182 Kalmar, Sweden}
\affiliation{Solid State Physics/The Nanometer Structure Consortium, Lund University, Box 118, SE-221 00 Lund Sweden}
\author{M. Fhokrul Islam}
\author{C. M. Canali}
\affiliation{Department of Physics and Electrical Engineering,
Linnaeus University, SE-39182 Kalmar, Sweden}
\author{M. R. Pederson}
\affiliation{Office of Basic Energy Sciences, SC22.1, U.S. Department of Energy, 1000 Independence Ave. SW Washington, DC 20585-1290}
\date{\today}

\begin{abstract}
Using first-principles methods we study theoretically the
properties of an individual $\{Fe_4\}$ single-molecule magnet (SMM) attached
to metallic leads in a single-electron transistor geometry.
We show that the conductive leads do not affect the spin ordering and
magnetic anisotropy of the neutral SMM.
On the other hand, the leads have a strong
effect on the anisotropy of the charged states of the molecule, which
are probed in Coulomb blockade transport.
Furthermore, we demonstrate that an external
electric potential, modeling a gate electrode, can be used to
manipulate the magnetic properties of the system.
For a charged molecule, by localizing the extra charge with the gate voltage
closer to the magnetic core,
the anisotropy
magnitude and spin ordering converges to the values found for the isolated $\{Fe_4\}$ SMM.
We compare these findings with the results of recent quantum transport
experiments in three-terminal devices.

\end{abstract}
\maketitle

\section{introduction}
\label{intro}

In recent years molecular spintronics has emerged as one of the most
active areas of research within magnetism at the atomic scale.\cite{rocha05,rocha06,trif10,sessoli99,bogani08,candini11}
Progress in the field is driven in part by advances
in chemical design and synthesis, which allow the realization of
interesting magnetic molecules
with desired electronic and magnetic properties. A
second essential feature of ongoing research is the improved
ability of integrating individual
magnetic molecules into solid-state nano-electronic devices.

Typically magnetic molecules have
long spin-relaxation times, which
can be utilized in
high-density information storage.
They are also usually characterized
by a weak hyperfine interaction with the environment,
resulting in long spin coherence times, which is an essential property for
applications in quantum information processing.
Single-molecule magnets (SMMs) are a special class of spin-ordered and/or magnetically active
molecules characterized by a relatively high molecular spin and large magnetic
anisotropy energy.\cite{sessoli06}
The latter lifts the spin degeneracy
even at zero magnetic field,
and favors one particular alignment of the spin, making the molecule
a nanoscale magnet.

One of the goals of molecular spintronics is to address the magnetic states
of individual magnetic molecules with electric fields and electric currents.
In the last six years experimental efforts toward this goal have
considered different classes of magnetic molecules and strategies to
incorporate them into electric nano-circuits. A particularly interesting
direction focuses on quantum transport in a single-electron transistor (SET),
a three-terminal device where
a SMM bridges the nanogap between two conducting nano-leads, and can further
be electrically manipulated by the gate voltage of a third nearby electrode.
In the regime of weak coupling to the leads,
the electric
charge on the central SMM is quantized and can be controlled
by the external gate. When Coulomb blockade is lifted by either gate or bias
voltages, transport occurs via tunneling of single electrons in and out of
the SMM.
Therefore a study of transport in this geometry can provide detailed information
on the magnetic properties of individual SMMs, both when the molecule is
neutral and when it is electrically charged.

Early SET experiments on SMMs\cite{heersche06,jo06} focused
on the archetypal SMM $\{Mn_{12}\}$-acetate\cite{sessoli93,sessoli06}, characterized by
the ground-state spin $S=10$ and a large
magnetic anisotropy barrier of approximately 50 K.
Unfortunately these experiments and studies of self-assembled molecules
on gold surfaces\cite{sessoli08} have shown that the magnetic properties of
$\{Mn_{12}\}$ complexes are extremely fragile and easily disrupted
when the molecule is attached to metallic leads or surfaces.

More recently, another class of SMMs, namely the tetranuclear $\{Fe_4\}$
molecule, has emerged as a candidate in molecular
spintronics that does not suffer the drawbacks of
$\{Mn_{12}\}$.
The properties of $\{Fe_4\}$ in its neutral state are well
studied in the crystal phase\cite{accorsi_2006, sessoli_2009}, and include
a molecular ground-state spin $S=5$ and an intermediate magnetic anisotropy
barrier $\approx 15$ K.
In contrast to what happens with $\{Mn_{12}\}$, these magnetic
characteristics remain stable when the molecule is deposited on a gold
surface.\cite{sessoli_NM_2009, sessoli10}
Furthermore, its tripodal ligands are shown
to be advantageous for the preparation of single-molecule electronic devices.
Indeed, recent three-terminal
quantum transport experiments\cite{alexander10,vzant11,vzant12},
with $\{Fe_4\}$
as the central island of a SET, show that
this molecule behaves indeed as a nanoscale magnet,
even when it is wired to
metallic leads. The magnetic anisotropy is significantly affected by adding or
subtracting single charges to the molecule\cite{alexander10}, an operation that
can be performed reversibly with the gate voltage.
More refined techniques\cite{vzant12}
allow the extraction
of the magnetic anisotropy of the neutral and charged
molecule from the transport measurements with unprecedented accuracy.

In this paper we carry out density functional theory (DFT)
calculations to evaluate
the magnetic properties of a $\{Fe_4\}$ connected to gold electrodes
and under the effect of an external electric field representing a gate voltage.
The geometry considered here is supposed to model
the typical situation realized in
current SET experiments, although some details might be different.
The main aim of our work is to investigate theoretically how
the spin ordering and the magnetic anisotropy of $\{Fe_4\}$ are affected
by weak coupling to the leads, both when the molecule is in its neutral
state and when a single charge is added to or subtracted from the device.
A second important objective of the paper is a theoretical analysis of how
these magnetic properties can be modified and controlled  by means of an external
electric potential representing a gate electrode.

Although a full-fledged first-principles study of quantum transport
is beyond the scope of the
present paper, as we explain below, we believe that our analysis of the
charged states under the effect
of an external electric field is useful to develop methods to compute the
tunneling conductance within a master equation formalism.
Ref.~\onlinecite{carlo10}
introduced a DFT description
of the neutral and charged states of an {\it isolated}
$\{Mn_{12}\}$ SMM, which were then used
in a master equation formalism for quantum transport.
Here the coupling to the leads was treated with
phenomenological tunneling amplitudes taken form experiment.
In Ref.~\onlinecite{ferrer_prl2009} charge transport
was studied by means of a non-equilibrium Green's function approach, which
included the presence of metallic leads in the regime of strong coupling.
Transport calculations carried out in the same transport regime
but based on microscopic tight-binding models have also been considered.\cite{renani2012}
This approach has the drawback that charging
effects, essential for the description of SET experiments, cannot be
adequately incorporated.

In this sense the present paper is a further contribution to these early
attempts to use first-principles methods based on DFT to investigate
quantum transport in a SET with a SMM.
We are aware that the use of DFT can be problematic when it comes to describing
the electronic structure and Fermi level alignment of molecules
coupled to external electrodes,
particularly when charged states are involved.
Assessing the limitations of DFT in this
context is also a technical aim of the present paper. In this regard,
we discuss potential uncertainties in the relative level alignment of the electrode and molecular states.

The main findings of our analysis are the following. The SMM $\{Fe_4\}$ in its
neutral state is indeed quite robust against the presence of metallic
leads: both spin ordering and magnetic anisotropy are essentially identical to
the one of the isolated molecule. For the case of a charged molecule,
the effect of the leads is more complex. Our calculations show that the
extra charge tend to reside primarily on the ligands between
molecule and metallic
leads, and only minimally on the magnetic core. As a result the addition of electrons
affects the magnetism of molecule (specifically the magnetic anisotropy)
considerably less than when
the molecule is isolated. We find that an external gate voltage can be used to
localize the extra charge closer to the central magnetic core, and in this
case the magnetic characteristics of the device converge to the ones of the
isolated $\{Fe_4\}$ SMM.

The organization of this paper is as follows. In
Sec.~\ref{theory} we present an overview of the theoretical and
computational
approach employed in this work. The electronic and magnetic
properties of different charge states of the isolated $\{Fe_4\}$ SMM is
discussed in Sec.~\ref{isolated}. The effects of the metallic
leads on the properties of the molecule are discussed in
Sec.~\ref{leads}. In Sec.~\ref{efield} we discuss how a
confining electric potential affects the magnetic properties of the
molecule. In Sec.~\ref{expe} we compare the results of our calculations with
recent SET transport experiments.
Finally in Sec.~\ref{conclusion} we summarize our
results.

\section{Theoretical background}%
\label{theory}

\subsection{Spin Hamiltonian and the giant-spin model}
\label{giant_spin}
In a first approximation, the exchange interaction between the magnetic
ions of a molecular magnet can be described by an isotropic Heisenberg model
\begin{equation}
H = \sum_{ij} J_{ij}{\bf s}_i \cdot {\bf s}_j\;,
\label{heisenberg}
\end{equation}
where ${\bf s}_i$ is the spin of the magnetic ion $i$ and
the constants $J_{ij}$ describe the super-exchange coupling between
ions $i$ and $j$. Clearly, the validity of Eq.~(\ref{heisenberg})
relies crucially on the assumption that each magnetic ion is characterized
by a well-defined quantum spin, localized at the ion position.
There might be cases where the spin-polarization of the molecule
is delocalized, where
this assumption may break down. Such cases are especially probable when an excess tunneling electron is present.  Once the exchange constants are known,
the Hamiltonian can be diagonalized. Since $H$ is a sum of scalars in
spin space, it commutes with the total spin $\bf S$. Therefore the eigenvalues
of ${\bf S}^2$ and ${S}_z$ can be used to label the eigenstates of $H$.

In the case of single-molecule magnets, the ground state (GS) of
Eq.~(\ref{heisenberg}) is characterized by a relatively large spin $S$,
and it is separated by a fairly large energy
$\Delta J \equiv {\rm Max}_{\{ij\}} \big[ J_{ij}\big]$ from excited states
with different total spins.
Thus, at low energies $< \Delta J$
the magnetic molecule behaves effectively as an atom with
a relatively large spin $S$, known as {\it `giant spin'}. The approximation of
restricting to the lowest spin multiplet is known as {\it giant-spin model}.
According to Eq.~(\ref{heisenberg}), each spin multiplet is degenerate.
In the next section we will discuss how spin-orbit interaction lifts this degeneracy,
splitting the $2S+1$ states of the GS multiplet.

Note that the ground-state spin $S$ is not
always the maximum value $S_{\rm max}$ allowed
by Eq.~(\ref{heisenberg}). Due to the
presence of antiferromagnetic components, the most
common situation encountered in SMMs is an intermediate value
$1 < S < S_{\rm max}$,
which quasi-classically corresponds to a ferrimagnetic spin configuration.
Below we will show how all the physical quantities entering
in the spin Hamiltonian of Eq.~(\ref{heisenberg}) and the value of the giant spin $S$
can be calculated within DFT.

\subsection{Spin-orbit interaction and magnetic anisotropy barrier}
\label{anisotropy}

Spin-orbit interaction introduces terms to Eq.~(\ref{heisenberg}) that break
rotational invariance in spin space. Up to second-order perturbation theory, these
terms, besides anisotropic corrections to the Heisenberg model,
include the antisymmetric Dzyaloshinskii-Moriya spin exchange, and the single-ion magnetic anisotropy
$H_{ia} = - \sum_i ({\bf d}_i\cdot{\bf s}_i)^2 $. Because of these terms, the total spin
is no longer a good quantum number. Within the giant-spin model of SMMs, where
the isotropic exchange is the dominant magnetic interaction, the main
effect of the spin-orbit interaction is to lift the spin degeneracy of the GS multiplet.
To second-order perturbation theory, this can be described by the following
anisotropy Hamiltonian for the giant spin operator ${\bf S} = (S_x, S_y, S_z)$

\begin{equation}
{\cal H}=DS_z^2+E(S_x^2-S_y^2)\;.
\label{H_D}
\end{equation}
The parameters $D$ and $E$ specify the axial and transverse magnetic
anisotropy, respectively. If $D<0$ and $|D| >> |E|$, which are
defining properties for SMMs, the system exhibits an easy axis in
the $z$-direction. In the absence of magnetic field, and neglecting
the small transverse anisotropy term, the GS of Eq.~(\ref{H_D}) is
doubly degenerate and it corresponds to the eigenstates of $S_z$
with eigenvalues $\pm S$. To go from the state $S_z = +S$ to the
state $S_z = -S$ the system has to surmount a magnetic anisotropy
energy barrier $\Delta E = |D|S^2$. In addition, transitions which
change the axial quantum numbers require some type of carrier to
balance the change in spin. When the transverse term is not
negligible and $S_z$ is not a good quantum number, we can still
define an anisotropy barrier separating the two (degenerate) lowest
energy levels as the energy difference between GS energy and the
energy of the highest excited state. If $D>0$ the systems has a
quasi-easy plane perpendicular to the $z$-axis without energy
barrier.

The anisotropy parameters $D$ and $E$ can be calculated within a self-consistent-field (SCF)
one-particle theory (e.g. DFT or Hartree Fock),
by including the contribution of the spin-orbit interaction. Here we summarize the
main steps of the procedure originally introduced in Ref.~\onlinecite{mark99}.
(For more recent reviews see Refs.~\onlinecite{mark03, mark06}.)

The starting point are the matrix elements of the spin-orbit interaction (SOI) operator
\begin{equation}
U({\bf r}, {\bf p} , {\bf s})= -\frac{1}{2c^2} {\bf s}\cdot {\bf p}
\times {\boldsymbol \nabla} \Phi ({\bf r})
\end{equation}
($\bf p $ is the momentum operator;  $ \bf s$ is the electron spin operator;
$\Phi$ is the Coulomb potential and $c$ is the speed of light),
taken with respect to the unperturbed single-particle spinor wave functions
$|\psi_{k\sigma}\rangle =|\phi_{k\sigma}\rangle|\chi_{\sigma}\rangle$, which
are solutions of the SCF-approximation Schr\"odinger equation
\begin{equation}
H|\psi_{k\sigma}\rangle = \epsilon_{k\sigma} |\psi_{k\sigma}\rangle
\end{equation}
Here $\phi_{k\sigma}({\bf r}) \equiv  \langle {\bf r}||\phi_{k\sigma}\rangle$ is the
orbital part of the wavefunction; the two-component spinors
$|\chi_{\sigma}\rangle,\ \sigma = (1, 2)$ are the eigenstates of
${\bf s}\cdot {\bf \hat n}$, where the unit vector
${\bf \hat n} = {\bf \hat n}(\theta, \varphi)$
is an arbitrary quantization axis.

The matrix elements can be written as\cite{mark99}
\begin{eqnarray}
U_{k\sigma,k'\sigma'} &=& \left\langle\psi_{k\sigma}\right|
U ({\textbf r},{\textbf p},{\textbf s})  \left| \psi_{k'\sigma'} \right\rangle
\nonumber  \\
&=& -i \sum_i \left\langle \phi_{k\sigma}\right| V_i\left| \phi_{k'\sigma'} \right\rangle
 \left\langle \chi_{\sigma} \right| s_i \left| \chi_{\sigma'} \right\rangle\;,
\label{eq:matrixelements}
\end{eqnarray}
where the matrix elements of the operator ${\bf V} = (V_x, V_y, V_z) $ are defined as
\begin{eqnarray}
 &&\left\langle \phi_{k\sigma}\right| V_x\left| \phi_{k'\sigma'} \right\rangle =
\nonumber \\
&& \frac{1}{2c^2}\left(
\left \langle \frac{\partial \phi_{k\sigma}}{\partial z}\right|\Phi \left|
                       \frac{\partial \phi_{k'\sigma'}}{\partial y}\right\rangle
-\left \langle \frac{\partial \phi_{k\sigma}}{\partial y}\right|\Phi \left|
                       \frac{\partial \phi_{k'\sigma'}}{\partial z}\right \rangle
\right)\; ,
\label{V_op}
\end{eqnarray}
and cyclical. Note that Eq.~(\ref{V_op}) avoids the
time-consuming calculation of the gradient of the Coulomb potential, replacing it with the
calculation of the gradient of the
basis functions in which $\phi({\bf r})$ is expanded. The above representation of the spin-orbit interaction arises by an integration by parts
of the matrix element defined in Eq.~(\ref{V_op}). It is similar to the form of spin-orbit interaction that comes out of the Dirac equation.

In the absence of an external magnetic field, the first-order perturbation-theory correction
to the {\it total GS energy} cause by the SOI is zero because of time-reversal symmetry.
The second-order correction
is nonzero and can be written as

\begin{equation}
\Delta_2 =\sum_{\sigma\sigma'}
\sum_{i,j}M_{ij}^{\sigma\sigma'}s_{i}^{\sigma\sigma'}{s_{j}^{\sigma'\sigma}}\;,
\label{aniso}
\end{equation}
where
\begin{equation}
s_{i}^{\sigma\sigma'}\equiv \langle\chi_\sigma|s_i|\chi_{\sigma'}\rangle\;,
\label{spin_me}
\end{equation}
and
\begin{equation}
M^{\sigma\sigma'}_{ij} \equiv
-{\sum_{k={\rm occ}}}\ {\sum_{k'={\rm unocc}}}
\frac{\langle\phi_{k\sigma}|V_i|\phi_{k'\sigma'}\rangle
\langle\phi_{k'\sigma'}|V_j|\phi_{k\sigma}\rangle}{\epsilon_{k\sigma}-\epsilon_{k'\sigma'}}\;,
\label{tensor}
\end{equation}
where the sums over $k$ and $k'$ involve occupied and unoccupied states,
respectively.

Eq.~(\ref{aniso}) is the central equation in the study of the magnetic anisotropy. Since
the spin matrix elements
$s_{i}^{\sigma\sigma'}$ depend
on the orientation of the arbitrary axis of quantization ${\bf \hat n}$,
so does also the total energy shift. This is
precisely the origin of the magnetic anisotropy brought about by SOI.

We now consider the case of a closed-shell molecule, a system
with a well defined HOMO-LUMO gap in order to avoid problems with partial occupancy,
with $\Delta N$ excess of majority spin electrons.

We have the important relation
\begin{equation}
\left \langle\chi_1\right| s_i \left|\chi_1\right\rangle =
- \left \langle\chi_2\right| s_i \left|\chi_2\right\rangle =
\frac{\left\langle S_i\right\rangle}{\Delta N}
\label{espin_to_giant_spin}
\end{equation}
where $\left\langle S_i\right\rangle$ is the GS expectation value of the $i^{th}$-component
of the total spin of the system for the given choice of the quantization axis.
On the basis of our discussion of the giant-spin model, $\left\langle S_i\right\rangle$
can be re-interpreted as the expectation values of the components of the
giant-spin operator $\bf S$ for the
spin-coherent state $\left|S, {\bf \hat n}\right \rangle$ with $S= \Delta N /2$.

Using the resolution of the identity in spin space,
$\sum_{\sigma}\left |\chi_{\sigma}\right \rangle \left\langle \chi_{\sigma}\right|= 1$,
we can write
\begin{eqnarray}
&&\left \langle\chi_1\right| s_i \left|\chi_2\right\rangle
\left \langle\chi_2\right| s_j \left|\chi_1\right\rangle = \nonumber\\
&&\left \langle\chi_1\right| s_i s_j\left|\chi_1\right\rangle
-\left \langle\chi_1\right| s_i \left|\chi_1\right\rangle
\left \langle\chi_1\right| s_j \left|\chi_1\right\rangle\nonumber\\
&&= \left \langle\chi_1\right| s_i s_j\left|\chi_1\right\rangle
- \frac{\left \langle S_i\right\rangle\left \langle S_j\right\rangle}{(\Delta N)^2}
\label{off_diag_sisj}
\end{eqnarray}
and a similar expression for
$\left \langle\chi_2\right| s_i \left|\chi_1\right\rangle
\left \langle\chi_1\right| s_j \left|\chi_2\right\rangle$.

With the help of Eqs.~(\ref{espin_to_giant_spin}),  (\ref{off_diag_sisj}) and
$\left \langle\chi_{\sigma} \right| (s_i)^2 \left |\chi_{\sigma}\right\rangle = 1/4$, Eq.~(\ref{aniso}) takes the form
\begin{equation}
\Delta_2 = \alpha +
\sum_{ij} \gamma_{ij}
\left \langle S_i\right\rangle\left \langle S_j\right\rangle\;,
\end{equation}
where $\alpha = \sum_{ij} (M^{12}_{ii} + M^{21}_{ii})$ is a constant independent of the quantization axis.
The anisotropy tensor $\gamma_{ij}$ is given by
\begin{equation}
\gamma_{ij}=  \frac{1}{(\Delta N)^2}\sum_{ij} (M^{12}_{ij} + M^{22}_{ij} - M^{12}_{ij} - M^{21}_{ij})
\end{equation}
The tensor $\gamma_{ij}$ can now be diagonalized by a unitary transformation
and $\Delta_2$ becomes
\begin{eqnarray}
\Delta_2 &=& \alpha +
A (\left \langle S'_x\right\rangle)^2  +
B (\left \langle S'_y\right\rangle)^2  +
C (\left \langle S'_z\right\rangle)^2
\label{d2_diag1}\\
&=&  \alpha +
A \left \langle (S'_x)^2\right\rangle  +
B \left \langle (S'_y)^2\right\rangle +
C \left \langle (S'_z)^2\right\rangle\;,
\label{d2_diag2}
\end{eqnarray}
where $A, B, C$ are the eigenvalues of $\gamma_{ij}$ and
the $S'_i$ are the three spin components
rotated along its three principal
axis (Eq.~(\ref{d2_diag2}) follows from Eq.~(\ref{d2_diag1})
thanks to the properties of spin coherent states).

This expression for
$\Delta_2$
is exactly the expectation value
$\left\langle S, {\bf \hat n}\right| {\cal H}\left|S,{\bf \hat n}\right\rangle$
of the quantum spin
Hamiltonian
\begin{equation}
{\cal H} = \alpha + A (S'_x)^2 + B (S'_y)^2 + C (S'_z)^2\,,
\label{giant_spin2}
\end{equation}
in the spin coherent state $\left| S, {\bf \hat n}\right\rangle$.
Eq.~(\ref{giant_spin2}) is equivalent to Eq.~(\ref{H_D})
up to an irrelevant constant.

The perturbative method described here
works well for systems with a large HOMO-LUMO gap.
However, for systems that have nearly degenerate and not fully occupied HOMO
levels, which often is the case for charged molecules, the
perturbative approach breaks down, since some of the energy denominators in Eq.~(\ref{tensor}) vanish.
To avoid this problem
the magnetic anisotropy can alternatively be calculated by an exact diagonalization method.
In this approach, the solutions of the one-particle Schr\"odinger equation in the SCF approximation
(which does not include SOI),
are used to construct a  finite matrix representation of the SOI, which is then
diagonalized exactly. The matrix is then diagonalized subject to the constraint that the resulting spin is aligned along a give choice of the quantization axis. The resulting single-particle solutions
$\{\epsilon'_k, \left |\psi_k'\right\rangle = \sum_{\sigma}|\phi'_{k\sigma}\rangle |\chi_{\sigma}\rangle \}$
are used to compute the trace of the system as a function of direction of the
quantization axis (or average direction of the magnetization). In Ref.~\cite{hmm}, a
discussion of the relationship between the second-order variation in the trace and the self-consistent
second-order variation of total energy is presented.

Once one has obtained the trace as a function of axis of quantization,
one can use relatively standard techniques to decompose the trace into a
spherical harmonic representation and then determine the magnetic
principal axes. Alternatively by choosing magnetic principal axes that are equivalent to those predicted from the
second-order expressions, it is always possible to directly compare exact-diagonalization results with the second-order results.
Using exact diagonalization, One can further extract parts of the fourth- and higher-order anisotropy
terms as well. However, since self-consistency and other terms also affect the magnetic anisotropy
at fourth-order and beyond, the exact
diagonalization results are primarily used to determined whether the second-order results are
expected to be stable and a good approximation to experiment. In cases where near degeneracies occur at the
Fermi level, the second-order and exact-diagonalization results can be very different especially if the
states near the Fermi level are coupled by the spin-orbit interaction. For such cases, a much more careful
analysis of results is needed and it is reasonable to expect that some degree of self-consistency with
non-collinear capabilities will be needed.

For electronic-structure methods, such as NRLMOL, where the
wavefunctions are expanded in terms of atom-centered localized basis functions the second-order perturbative
method allows one to further analyze the anisotropy Hamiltonian on an atom-by-atom basis. By expanding
the Kohn-Sham orbitals ($\left| \phi_{k \sigma}\right\rangle$)
in Eq.~(\ref{tensor}) in terms of the atom-centered basis,
the second-order expression (Eq.~(\ref{aniso})) can be decomposed into a sum over four centers.~\cite{baruah} The
super-diagonal terms (all center indices the same) can then be used to determine an anisotropy Hamiltonian
associated with each atom. In Ref.~\cite{baruah}, this  decomposition has been used to verify
the perpendicular hard-axis alignment model in the Co$_4$ easy-plane magnetic molecule. In Ref.~\cite{park},
Baruah's method was used to demonstrate that essentially all of the magnetism in Mn$_{12}$-acetate was due to the
outer eight $S$=2 crown Mn ions, and that the sum of the single-ion anisotropies was very close to the
total anisotropy. Further the degree to which non-additivity occurred was explained by a canting of the atom
projected anisotropy axes relative to the global anisotropy axis.

To complete the discussion about second-order anisotropy Hamiltonians, derived either perturbatively or via
exact diagonalization, it is important to note a contribution to van Wullen and
coworkers.~\cite{vanwullen1,vanwullen2}. Van Wullen noted that once a method is used to determine the
spin-orbit energy as a function of axis of quantization that an additional quantum correction is needed
to determine the parameter's $D$ and $E$ in the anisotropy Hamiltonian. For example, the $M_s=0$ eigenstate is not
aligned with an axis of quantization along the $x$-axis, the $y$-axis or any other axis. Therefore more care must
be taken to determine $D$ once the classical energy as a function of expectation value of {\bf S} is known.
Accounting for this correction changes the definition of $D$, as originally derived by Pederson and Khanna, by a
factor of $(S+1/2)/S$. While this correction is small in the large $S$ limit, it can be important for systems
with lower spin.

\subsection{Computational details}
\label{computation}

In this paper we use a self-consistent field approximation based on
density-functional theory (DFT). A review of this approach in the study of
molecular magnets can be found in Ref.~\cite{mark06}.
Here we remind a few key features that are relevant for the present work.

In a DFT calculation of a magnetic molecule, we obtain
the total energy of the system for
specially prepared spin symmetry-breaking metastable states. In many cases these
are ferrimagnetic spin configurations, suggested by experiment. The energies of these
different metastable states can then be compared and the lowest-energy spin configuration
determined.
Alternatively, it is also possible to impose a fixed spin configuration, which in
principle would not remain stable after convergence. In all these symmetry-breaking
calculations the state with a given spin configuration is represented
by a single Slater determinant of occupied single-particle states,
constructed in terms of self-consistent
Kohn-Sham eigenvectors. In the absence of SOI, the Kohn-Sham wavefunctions
have a well-defined
spin character, majority or minority spin.
Therefore, the single Slater determinant, representing a given
spin configuration, is an eigenstate of the component of the total spin
 ${\bf S}$ in the direction
of the quantization axis $\bf \hat n$,
which is the magnetization direction. In general this single Slater
determinant is not an eigenstate of ${\bf S}^2$, but in many cases it will
have a large overlap with
the eigenstate of ${\bf S}^2$ with $S$ equal
to the eigenvalue of ${\bf S}\cdot {\bf \hat n}$.
The GS total spin $S$ of the molecule
(in the absence of SOI) is taken to be equal to one half of
the excess of majority spin
electrons,
$S= \Delta N/2 = (N_{\rm maj} - N_{\rm min})/2 $, for the metastable spin configuration
with the lowest total energy. The spin magnetic moment of the system
in units of the Bohr magneton $\mu_{\rm B}$
is then
$\mu_S = \Delta N\mu_{\rm B} = 2S \mu_{\rm B}$.
In the DFT study of SMMs the possible presence of fractional occupancy
of some of the KS wavefunctions
close to the Fermi energy
might result in noninteger values of $N_{\rm maj} - N_{\rm min} $.
Typically this happens when the HOMO-LUMO gap is very small or vanishing.
We will encounter
examples of this difficulty in the study of the charged states of $\{Fe_4\}$.
The existence of this
general problem
was first discussed by Janak {\em et al}~\cite{janak} in reference
to near degeneracies between $3d$ and $4s$
electrons in neutral isolated atoms.
In Ref.~\cite{pedjak91} a set of equations are derived which,
while cumbersome to solve, allow one
to variationally determine the electronic occupations that satisfy
the conditions proposed
by Janak in Ref.~\cite{janak}.

DFT can be used to extract the parameters defining the spin Hamiltonian that is supposed
to describe the exchange interaction between the magnetic ions of the molecule.
First of all DFT can be used to ascertain whether or not
there is a localized spin at each magnetic ion, by calculating the total spin polarization
inside a sphere centered about a given atom.
For typical SMMs, including the one considered
in this paper, while the magnetization density is not localized entirely on the magnetic
ions, the assumption of a well-defined quantum spin often turns out
to be quite reasonable.
Once this is established, the calculation of
the total energy for a few spin configurations
permits the computation of the exchange constants of Eq.~(\ref{heisenberg}).
We will see an example of this
in the next section.

The DFT calculations discussed herein
are performed with the Gaussian-orbital-based NRLMOL
program.\cite{mark90_1,mark90_2} All calculations employ the
Perdew-Burke-Ernzerhof\cite{perdew} (PBE) generalized-gradient
approximation for the density functional. A large basis set is
employed in which the exponents for the single Gaussians have been
fully optimized for DFT calculations. The NRLMOL code employs a
variational mesh for numerically precise integration and an analytic
solution of Poisson's equation.

All-electron calculations are performed for all elements of the
$\{Fe_4\}$ SMM except for the Au atoms that are used to construct the
leads attached to the molecule, for which we have used pseudo
potentials. All the electronic and magnetic properties are calculated
using an optimized geometry.

\section{Electronic and magnetic properties of isolated $\{Fe_4\}$ SMM}
\label{isolated}

\begin{figure}[h]
{\resizebox{2.5in}{2.5in}{\includegraphics{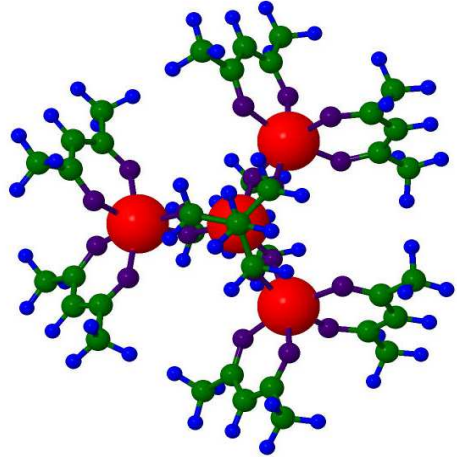}}}
\caption{(Color online) Ball-stick top view of an isolated $\{Fe_4\}$ SMM. Red, green, blue and purple balls correspond to iron, carbon, hydrogen and oxygen atoms, respectively}
\label{Fe4}
\end{figure}

The chemical composition of the molecule used in this work is
$Fe_4C_{76}H_{132}O_{18}$.\cite{accors06} The four Fe atoms in
$\{Fe_4\}$ SMM form an equilateral triangle, as shown in
Fig.~\ref{Fe4}. The molecule has idealized $D_3$ symmetry with the $C_2$
axis passing through the central atom and one of the peripheral
atoms. Using first-principles methods we have calculated, in detail,
the electronic and the magnetic properties of the $\{Fe_4\}$ SMM.

\begin{figure}[h]
\includegraphics[scale=0.35]{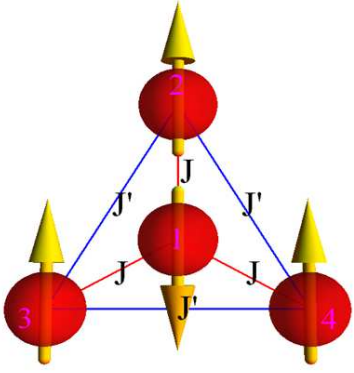}
\caption{Exchange interaction constants  between four Fe atoms in $\{Fe_4\}$
         SMM.}
\label{Fe4_J}
\end{figure}

In the ground state, the central Fe atom is coupled anti-ferromagnetically
with three peripheral atoms, whereas the three peripheral atoms
are coupled ferromagnetically with each other, as shown in
Fig.~\ref{Fe4_J}. Each of the four Fe atoms has spin $S_{Fe}=5/2$,
thus the total spin of the ground state is $S=5$. The magnetic
interactions between these atoms can be described by the Heisenberg spin
Hamiltonian (Eq. (\ref{H_D})),

\begin{eqnarray}
H&=&J({\bf s}_1\cdot{\bf s}_2+{\bf s}_1\cdot{\bf s}_3 +{\bf
s}_1\cdot{\bf s}_4) \nonumber \\
 && \ \ \ \  + J'({\bf s}_2\cdot{\bf s}_3+{\bf s}_3\cdot{\bf
s}_4+{\bf s}_4\cdot{\bf s}_2)\;.
\label{ham}
\end{eqnarray}

The two exchange parameters $J$ and $J'$ can be written in terms of
the expectation values of the Hamiltonian of Eq.~(\ref{ham}), for
different spin configurations.
\begin{eqnarray}
J=-\frac{2}{75}(E_{duuu}-E_{uuuu})\;, \nonumber \\
J'=\frac{1}{75}(2E_{duuu}-3E_{uudd}+E_{uuuu})\;.
\label{J}
\end{eqnarray}
Here, $E_{duuu}$, $E_{dduu}$ and $E_{uuuu}$ are the energies of the
molecule where the spin orientations at their respective atomic
positions (1,2,3,4) are labeled as $d={\rm down}$ or $u= {\rm up}$. Using NRLMOL
we have calculated the energies of different spin configuration and
upon substitution in Eq.~(\ref{ham}), we obtain $J=9.94$ meV and
$J'=0.64$ meV. DFT calculations overestimates $J$ and $J'$ since
estimated values from susceptibility measurements\cite{barra99} are
2.62 meV and 0.14 meV, respectively. However, we note that the ratio
of these two parameters agrees quite well for both theory and
experiment. These values of the exchange constants ensure that the
GS of the spin Hamiltonian Eq.~(\ref{J}) has indeed a total spin
$S = 5/2$, well separated
from excited states characterized by other values of $S$.

\begin{figure}[h]
{\resizebox{3in}{2.2in}{\includegraphics{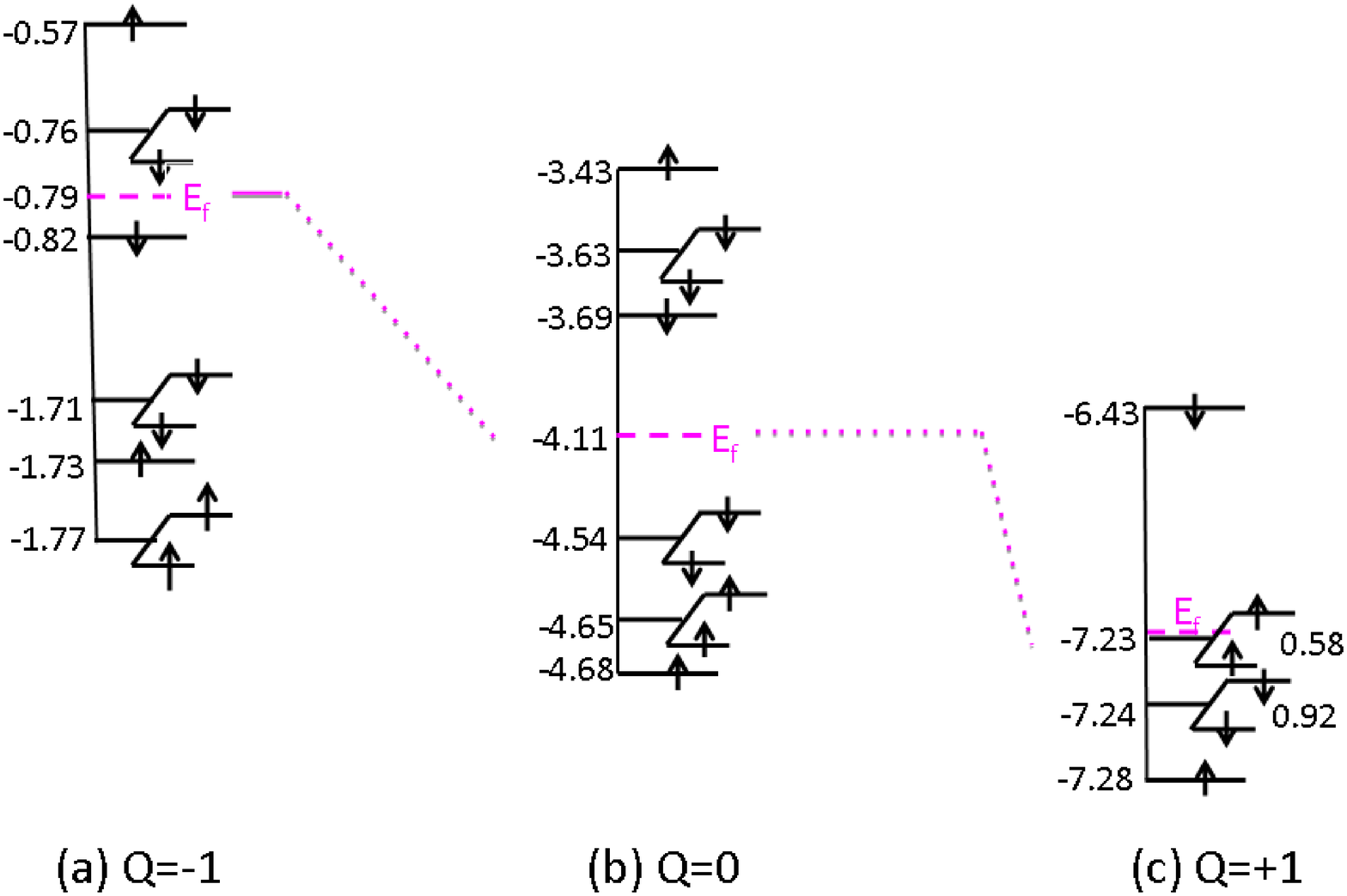}}}
\caption{Energy levels (eV) of different charge states of isolated
         the $\{Fe_4\}$ SMM without spin-orbit coupling. $E_f$ represents
         the Fermi level. The numbers on the right of the
         HOMO and HOMO-1 levels in (c) are the fractional occupancies of the
         corresponding level.}
\label{elevels_iso}
\end{figure}

Using first-principles methods we have calculated the electronic and
magnetic properties of $\{Fe_4\}$ SMM for the neutral ($Q=0$) and two
charged states, namely, the anion ($Q=-1$) and the cation ($Q=+1$).
(We will refer to the value $Q$ as the {\it extra charge} added to the system.)
A summary of the results is shown in Table~\ref{ch:iso}.
These results can be understood with the help of the structure of the
single-particle energy levels around the Fermi level in the absence of SOI,
plotted in Fig.~\ref{elevels_iso} for the three charge states $Q= 0, \pm 1$.

\begin{table}[h]
\caption{Properties of the isolated $\{Fe_4\}$ SMM for the three different charge states. *Note that
the energy gap reported for the cation refers to the energy difference between the half-filled HOMO and
empty LUMO. See Fig.~\ref{elevels_iso}(c).}
\label{ch:iso}
\begin{tabular}{|c|c|c|c|} \hline
Charge & Spin magnetic   & HOMO-LUMO  & Anisotropy \\
state  & moment $\mu_S$($\mu_B$)&energy gap (eV) &  barrier (K)      \\ \hline
Q=0    &  10.0         &  0.85      &  16.05     \\ \hline
Q=+1   &   9.3         &  0.80$^*$   &  53.42     \\ \hline
Q=-1   &   9.0         &  0.06      &   1.88     \\ \hline
\end{tabular}
\end{table}

The neutral molecule has a stable $S=5/2$ GS, as anticipated.
The  HOMO-LUMO  gap of the neutral isolated {$Fe_4$} molecule is  is about 0.85 eV, where both
HOMO and LUMO levels are minority (down) spin states, see Fig. \ref{elevels_iso}(b).
Apart from a small swapping of two levels below the HOMO, the
level structure for the $Q=-1$
charge state can be obtained from the the energy levels of the neutral system
simply by filling the neutral LUMO with a spin-down electron, see Fig. \ref{elevels_iso}(a). As a result
the total spin of the anion is $S= 9/2$.
Note also that
the HOMO-LUMO gap of the anion is now reduced to 0.06 eV
compared to the neutral molecule. The
electronic states changes significantly for the $Q=+1$ charge (cation) state: the two
doubly degenerate spin-up HOMO-1 and spin-down HOMO states of the neutral molecule
swap place, see Fig. \ref{elevels_iso}(c). Furthermore, since there is now one fewer electron, the new HOMO
is now half-filled. This implies that the total spin of the cation is again reduced
with respect to the neutral molecule by 1/2, that is, $S= 9/2$. The fractional occupancy
of the HOMO
plays important role in the enhancement of the magnetic
anisotropy, discussed below.
Our DFT calculations yield values of the total spin $S$ in agreement with the
level structure of Fig.~\ref{elevels_iso}.
In particular for the charged states $Q= \pm 1$, even when
the initial spin configuration is set to be $S= 11/2$, the system converges eventually
to the value $S= 9/2$.

For the neutral molecule, the magnetic anisotropy landscape is characterized by
an easy axis in
the direction perpendicular to the plane containing the four Fe
atoms (the $z$-axis).
As shown in  Table~\ref{ch:iso}, we find that the anisotropy barrier for this case
is about 16 K, which is in agreement with previous
calculations\cite{mark05} (In Ref \cite{mark05}, the authors have
used two different $\{Fe_4\}$ complexes. The molecular symmetry for
one of these complexes is $C_2$, whereas the the other one has $D_3$
symmetry. Our calculations agree well with the one that has $D_3$
symmetry.). Note that a well-defined energy gap between occupied and
unoccupied states (regardless of the spin),
ensures that the perturbative and exact
calculation of the anisotropy coincide.

The magnetic anisotropy of the $Q=+1$ charge state has
also uniaxial character in the
$z$-direction, with a barrier of about 53 K,
significantly larger than that obtained for the neutral
molecule. On the other hand for the $Q=-1$ charge state
the anisotropy is reduced to
about 1.9 K, with an easy axis in the $XY$-plane of the Fe atoms.
The large change in the anisotropy for the two charged
states has a very different
origin for the two cases and can be understood in the following way.

For the $Q=-1$ case, the small gap between the like-spin HOMO and LUMO states
might at first suggest a breakdown of the perturbative treatment.
In fact, these
two states are coupled only minimally by SOI. The most important coupling
occurs between the spin-down HOMO and the spin-up LUMO+1 states.
The energy difference
between these two states is $\approx 0.25$ eV. This value and the
corresponding energy denominator in Eq.~(\ref{tensor}) is large enough for
perturbation theory to work
(as a comparison with the exact approach clearly shows) and,
at the same time, small enough for this individual transition
to completely determine the
main features of the anisotropy landscape. In particular, it turns out that
this term in Eq.~(\ref{tensor}) favors an easy axis along
a direction in the $XY$-plane
of the Fe atoms. Since this magnetization direction is unfavorable for
other terms in  Eq.~(\ref{tensor}) (which prefer the $z$-direction), there are
positive and negative contributions in the total energy difference for the two
magnetization directions (calculated with Eq.~(\ref{aniso})),
which in the end lead to a reduced anisotropy
barrier.

The large enhancement of the magnetic anisotropy barrier found for the
$Q=+1$ state can also be understood with the help of
single-particle energy diagram
shown in Fig.~\ref{elevels_iso}(c). We observe that the half-filled
doubly-degenerate spin-up HOMO level lies just above a (close-to-100\%)
filled doubly
degenerate spin-down HOMO-1 level. The term involving transitions between
these two occupied and unoccupied levels totally
dominate Eq.(\ref{tensor}). In fact, the smallness
of the corresponding energy denominator (a few meV) renders
the perturbative approach
inadequate. This is a classical example of a quasi-degeneracy at the Fermi
level, where the exact treatment of SOI is necessary.
As it is often the case,
the inclusion of SOI lifts the quasi-degeneracy for a particular
direction of the magnetization, leading to a substantial decrease
of the total energy for that direction and, consequently,
to a large anisotropy energy
barrier.

\section{$\{Fe_4\}$ SMM attached to metallic leads}
\label{leads}

\begin{figure}[h]
{\resizebox{3in}{2.2in}{\includegraphics{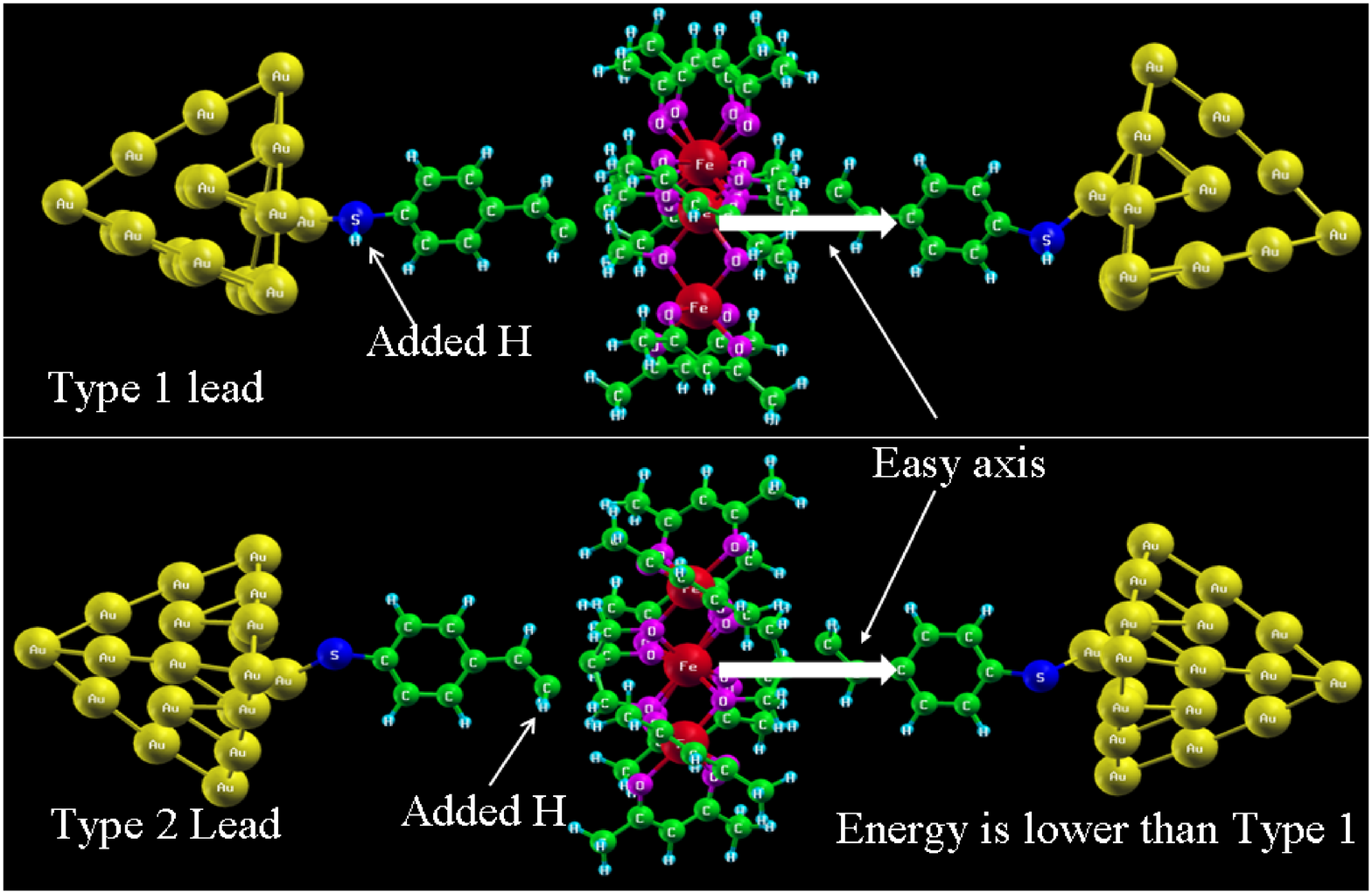}}}
\caption{$\{Fe_4\}$ SMM connected to Au$_{20}$ leads. Two types of leads are
          used in these calculations. In Type-1 lead a H atom is added to the S
          atom near the gold lead (top figure). In Type-2 lead a H atom is added to a
          C atom in the phenyl group near the $\{Fe_4\}$ molecule (bottom figure).}
\label{lead}
\end{figure}

In this section we investigate how the electronic and magnetic
properties of the $\{Fe_4\}$ SMM change when the molecule is attached
to metallic leads, as in transport experiments. The system that
we have in mind is a single-electron transistor device, where metallic
nano-leads, separated by a nanogap created by either break junction or
electric migration, are bridged by a molecule functionalized with
convenient chemical ligands. In our theoretical modeling, we are forced to
find a convenient finite representation of otherwise semi-infinite leads
in the form of finite clusters. For the calculations reported in this paper
we have chosen to model a metallic nano-lead with a finite
cluster of 20 gold ($^{79}$Au) atoms, arranged in a special
tetrahedral structure,
which can be viewed as a fragment of the face-centered cubic lattice of bulk Au
(see Fig.~\ref{lead}). This metal cluster, Au$_{20}$, has been previously investigated by
by Li {\it et al.} in Ref.~\onlinecite{li03}, where it was shown
that 20 Au atoms arranged in this geometric configuration form a very stable
system. Its very large HOMO-LUMO gap (1.77 eV) makes Au$_{20}$
chemically very inert. What is also important is that
its unique tetrahedral structure makes this cluster an ideal model for
Au surfaces at the nanoscale.
Our rational for using this cluster to model
metal leads is that during the fabrication of the nanometer-spaced electrodes,
via self-breaking by electro-migration for
example, the ensuing Au nano-leads will relax into
the most stable configuration which might be
well described by tetrahedral Au$_{20}$.
In a way, the tetrahedral Au$_{20}$ is the best representation of
bulk Au at the nanoscale.

Secondly, since ultimately we would like to investigate
transport properties of this system in coulomb blockade regime, it
is essential that coupling between the leads and molecule is weak.

After constructing the two leads in the form of two  Au$_{20}$ clusters,
we have connected the molecule via
phenyl groups, as shown in Fig.~\ref{lead}.
The functionalization of the ligands of
$\{Fe_4\}$ SMM by means of phenyl groups is
well-known and suitable to attach the molecule to Au surfaces \cite{sessoli10,accorsi_2006}.

A similar way of connecting the $\{Fe_4\}$ SMM to Au leads is employed in
SET experiments in the Coulomb blockade
regime\cite{alexander10,vzant11,vzant12},
and it ensures that the electronic
coupling between the molecule
and electrodes is weak.
To maintain proper bonding
and ionic neutrality of the entire cluster we have further removed one
hydrogen (H) atom from the molecule close to the contact point and
have added it to the chain part of the cluster. We have considered two
different ways of doing this. In the first case
(hereafter called Type-1 lead) we
have added the H atom to the sulfur (S) atom near the Au cluster.
In the second case (hereafter called Type-2 lead) we have added
H to the carbon (C) atom near
the contact point as shown in Fig.~\ref{lead}. After connecting the
leads with the molecule we have relaxed the entire system again.
Typically we find that the system with Type-2 leads is more stable,
that is, its energy is approximately 0.6 eV lower than the energy of the
system with Type-1 leads. We will nevertheless report results for both
cases unless otherwise specified.

A summary of the magnetic properties of the neutral molecule attached to
A$_{20}$ leads is shown
in Table~\ref{tb:lead}.
We can see that the combined molecule  plus leads system maintains
a sizable HOMO-LUMO gap of about 0.75 eV.
The first important result is that coupling the leads
does not cause a change of the spin of the molecule, which remains
equal to the value of the isolated neutral $\{Fe_4\}$, $S= 5$.
Secondly, the leads have
a very small effect also on the
magnetic anisotropy of the system: the magnetic anisotropy landscape
has still an easy axis
along the same
$z$-direction (see white arrow in Fig.~\ref{lead}),
with an energy barrier
quite close to the 16 K of the isolated molecule.

\begin{table}[h]
\caption{Properties of neutral $\{Fe_4\}$ SMM attached to A$_{20}$ leads,
compared to the properties of the isolated molecule (first row).
Type-1 and Type 2 (called in the paper also Type-1 lead  and Type-2 lead) refer to the two different choices to place an Hydrogen atom to the ligand, See
Fig.~\ref{lead}
}
\label{tb:lead}
\begin{tabular}{|c|c|c|c|} \hline
                  & Spin magnetic & HOMO-LUMO  & Anisotropy \\
                  & moment $\mu_S$($\mu_B$)& gap (eV)      &   barrier (K)      \\ \hline
$\{Fe_4\}$        &   10          &  0.85      &  16.05     \\ \hline
$\{Fe_4\}$+Type 1 &   10          &  0.87      &  15.99     \\ \hline
$\{Fe_4\}$+Type 2 &   10          &  0.57      &  15.47     \\ \hline
\end{tabular}
\end{table}

%
%
%
%
%
%
\begin{figure}[h]
{\resizebox{3in}{2.2in}{\includegraphics{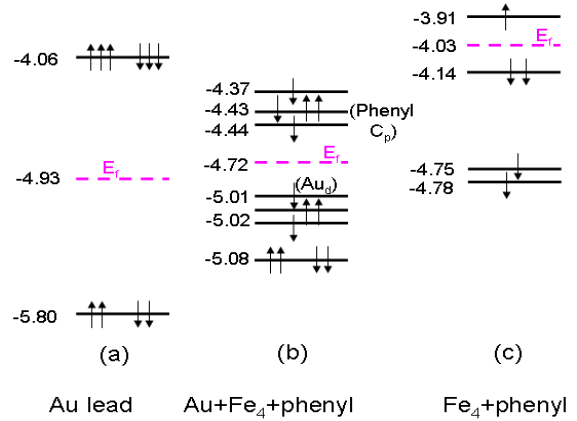}}}
\caption{Shifts in the energy levels (eV) of neutral $\{Fe_4\}$ SMM when connected to two Au$_{20}$
         leads (of Type-2). (a) Isolated Au lead (b) Leads+$\{Fe_4\}$+phenyl and (c)
         $\{Fe_4\}$+phenyl. The labels Au$_d$ and Phenyl C$_p$ in (b) indicate that
         the main contribution to those levels comes from $d$ levels of the Au leads and $p$
         levels of C in the phenyl ligands respectively.}
\label{elevel_lead}
\end{figure}
\begin{figure}[h]
\subfigure[Homo-neutral]{\includegraphics[scale=0.20]{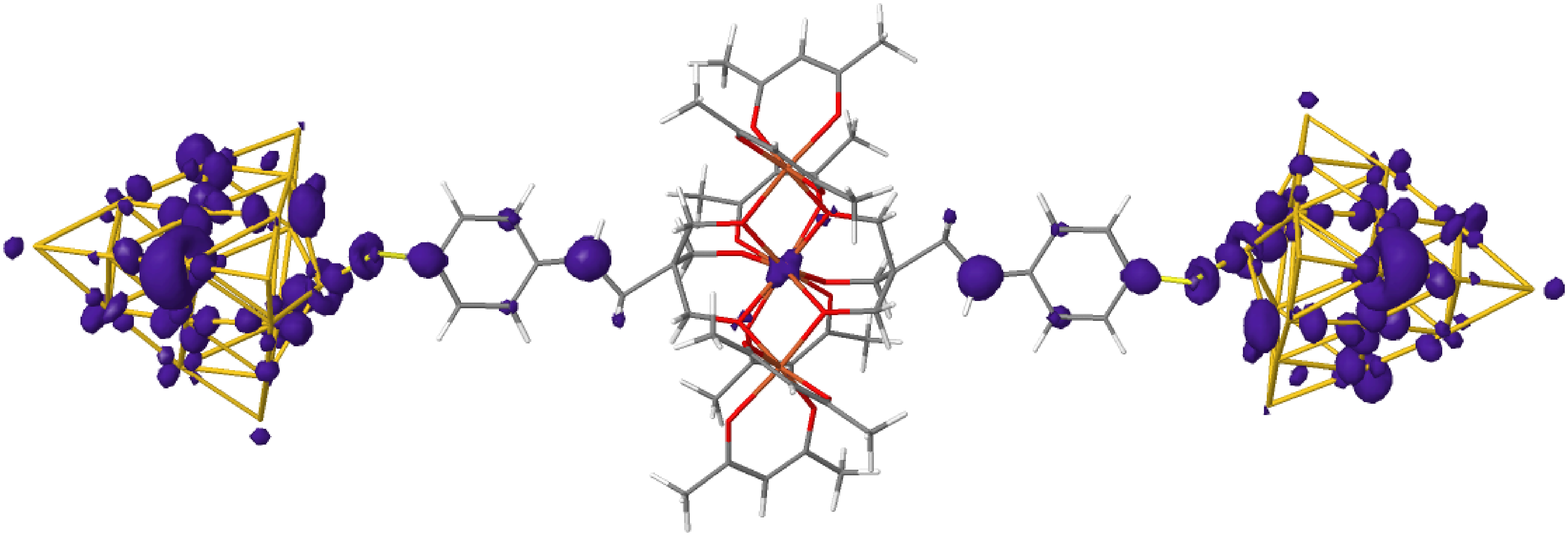}}
\subfigure[Lumo-neutral]{\includegraphics[scale=0.20]{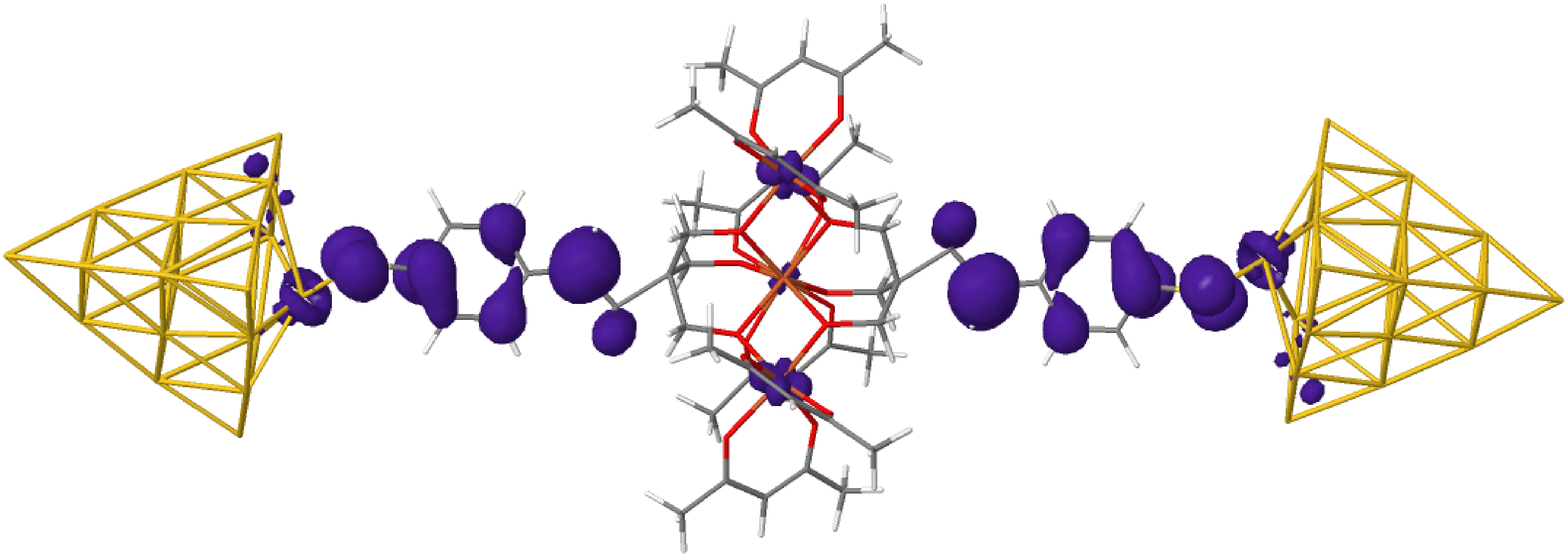}}
\caption{HOMO and LUMO of the neutral molecule.}
\label{neutral_homo_lumo}
\end{figure}
We can gain some insight about the robustness of the magnetic
structure of $\{Fe_4\}$ SMM under the influence of metallic contacts
by investigating the changes in the single-particle energy levels and level alignment and
occurring when the
leads are connected to the molecule. We have calculated separately the energy
levels of the isolated Au lead(s) and the $\{Fe_4\}$ + phenyl group, along with the levels of the
combined system [Au leads +  $\{Fe_4\}$ + phenyl group]. The results are shown in Fig.~\ref{elevel_lead}.
The states at and near the Fermi level of the two subsystems are dominated by
the $d$-levels of Au atoms and the $p$ levels of the
C atoms of the phenyl group. Thus, when the two systems are
combined, the charge transfer taking place to align the Fermi energies of the two subsystems
is restricted only within the contact
region, leaving the magnetic properties of the  $\{Fe_4\}$ inner core unchanged.
Fig.~\ref{elevel_lead}(b) shows that the energy levels of the combined system
around at the Fermi level correspond primarily to states
remote from the magnetic core.
This is also supported by Fig.~\ref{neutral_homo_lumo},
where we plot the probability density for the
HOMO and LUMO states of the $\{Fe_4\}$ + leads system. Both states have negligible contributions
on Fe atoms or atoms immediately nearby to these.
As we will see below, this implies that the magnetic properties
will remain unchanged even when extra charge is added to (or subtracted form) the system.

\begin{figure}[h]
{\resizebox{3.3in}{2.6in}{\includegraphics{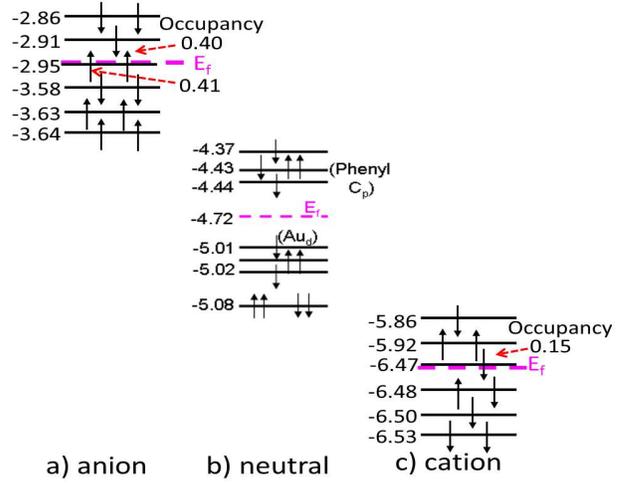}}}
\caption{Energy levels (eV) for the three charge states $Q= 0, \pm 1$ for the $\{Fe_4\}$ SMM
         connected to two Au$_{20}$
         leads (of Type-2). (a) Anion. (b) Neutral (same as in Fig.~\ref{elevel_lead}(b)). (c) Cation.
         The labels Au$_d$ and Phenyl C$_p$ in (b) indicate that
         the main contribution to those levels comes from $d$ levels of the Au leads and $p$
         levels of C in the phenyl ligands respectively.}
\label{charged_elevel_lead}
\end{figure}

Fig.~\ref{elevel_lead}(b) shows that both the non-degenerate spin-down HOMO
and the non-degenerate spin-down LUMO of the neutral
system lie quite close in energy to degenerate levels (occupied and non-occupied respectively).
Therefore we can expect that subtle energy-level
swaps may occur when one electron is added to or subtracted from the system.
As it is evident from Fig.~\ref{charged_elevel_lead}, this is exactly
what the calculations show. For the case of leads of Type 2, the HOMO of the anion
($Q= -1$) is now a half-occupied
doubly-degenerate spin-up level, lying very close to a non-degenerate spin-down LUMO.
This leads to a GS spin $S= 11/2$, and to a spin magnetic moment
close to $\mu_s = 11 \mu_{\rm B}$ (See Table \ref{ch:lead}).\footnote{The fact that
in Table \ref{ch:lead} the spin magnetic moment
of the $Q= -1$ charge state is not exactly equal to an integer is due to the fractional occupancy
(approximately 15\%) of the spin-down LUMO in our calculations, carried out with a finite smearing of
the Fermi-Dirac distribution.} A similar situation occurs for the $Q= +1$ charge state, which has
a GS spin $S=11/2$. We find that this state is however almost degenerate with another state
with $S= 9/2$.
For leads of Type-1 (which are less stable),
the spin configurations $S=11/2$  and $S= 9/2$ are almost degenerate
for both charged states, $Q= \pm 1$. Note that the spin magnetic moment of the $Q$= -1 charge state
is now closer
to  $\mu_S= 9\mu_{\rm B}$.
The quasi-degeneracy of two different spin configurations is a situation
where the assumption of the existence of a well-defined giant-spin model
may not be entirely adequate.

%
%
%
%
%
%
\begin{table}[h]
\caption{Magnetic properties of the three charge states when the $\{Fe_4\}$ SMM is attached
to Au leads as in Fig.~\ref{lead}.}
\label{ch:lead}
\begin{tabular}{|c|c|c|c|c|}   \hline
Charge & \multicolumn{2}{c|}{Spin magnetic } & \multicolumn{2}{c|}{Anisotropy } \\ 
state  & \multicolumn{2}{c|}{moment   $\mu_S$($\mu_B$)}  &  \multicolumn{2}{c|}{barrier (K)}  \\ \hline
& Type 1   & Type 2 & Type 1 & Type 2  \\ \hline
Q=0    & 10.0     & 10.0   & 15.99  & 15.47   \\ \hline
Q=+1   &  9.0     & 10.95  & 17.73  & 14.74   \\ \hline
Q=-1   &  9.6     & 10.65   & 11.23  & 16.97   \\ \hline
\end{tabular}
\end{table}
%

%


%
%
%
%
\begin{figure}[h]
\subfigure[ ]{\includegraphics[scale=0.20]{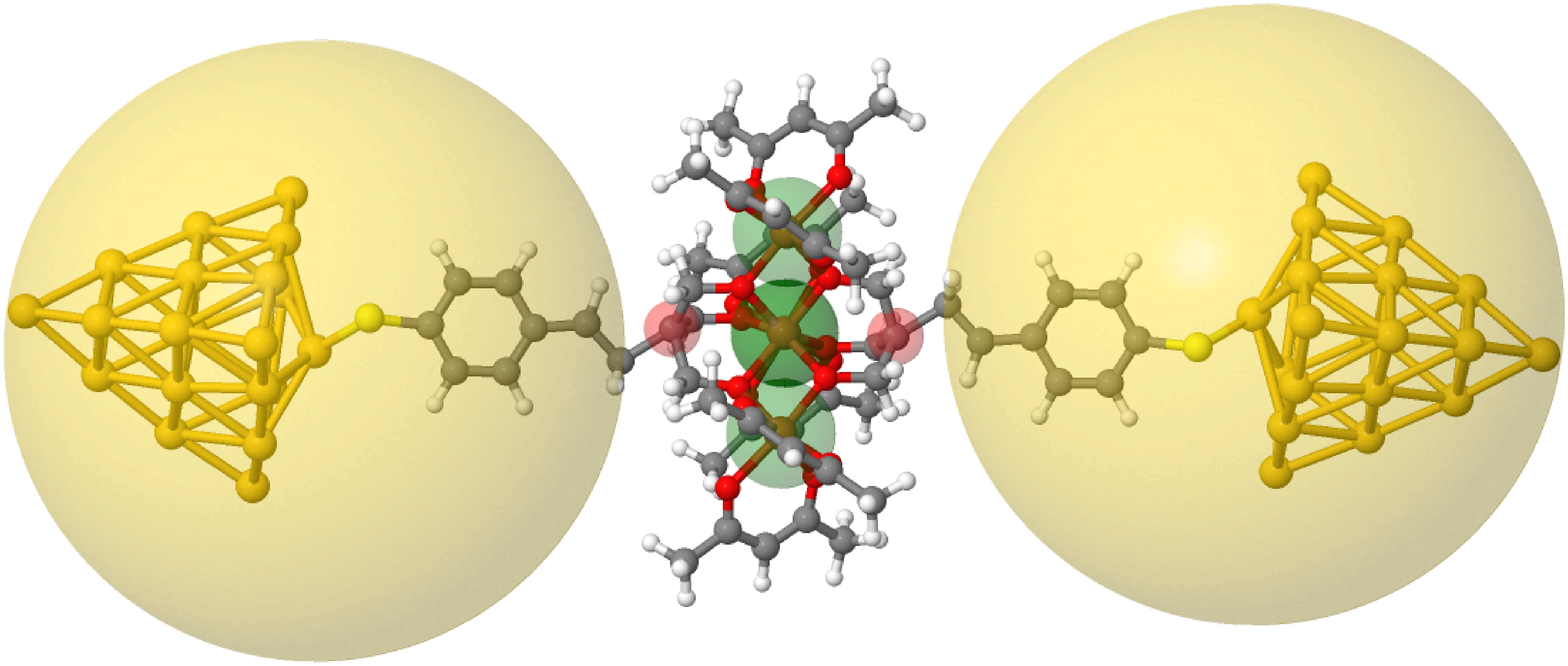}\label{fig:sA}}
\subfigure[ ]{\includegraphics[scale=0.21]{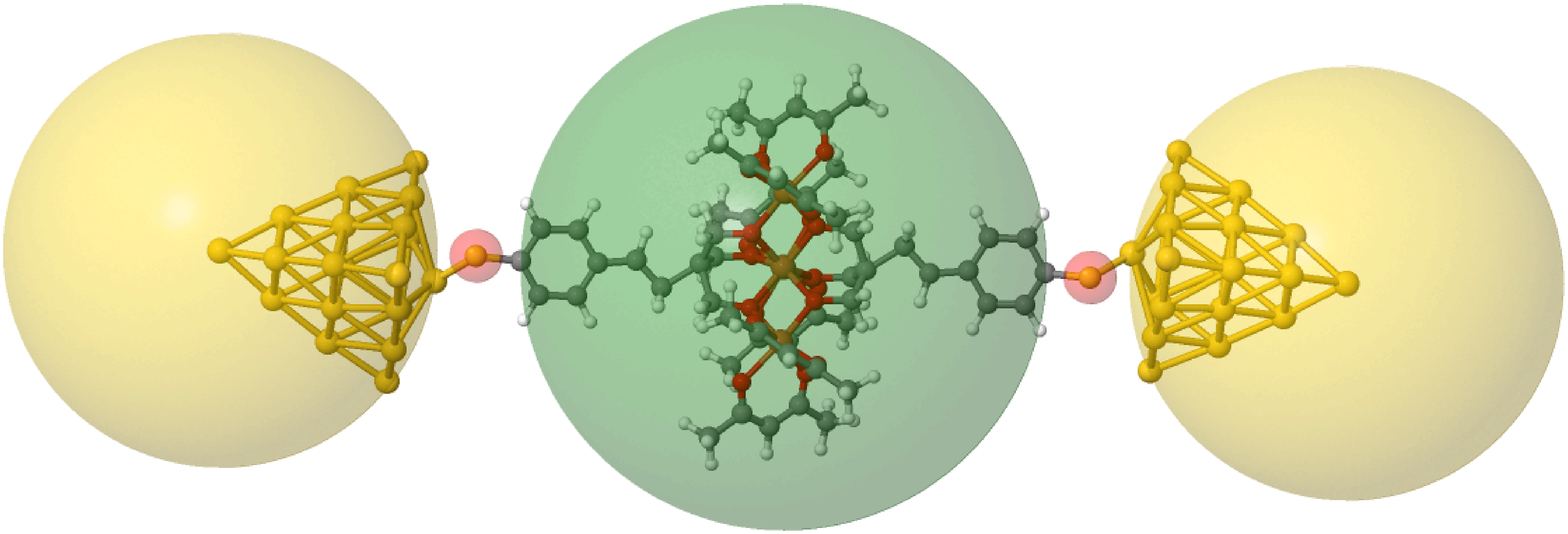}\label{fig:sB}}
\caption{(Color online) Evaluation of the fraction of
extra charge for the anion state ($Q= -1$, one extra electron added) with respect to the neutral state,
contained in different regions of the $\{  Fe_4\}$
molecule plus leads and phenyl groups. In (a) each yellow sphere surrounding
the lead and the phenyl group
contains
about 40\% of an electron charge.
In (b) each yellow sphere,  surrounding only the Au lead,
contains 21\% of electron charge. Therefore, the amount of charge
transferred to the leads is about 42\% and to the phenyl groups is 38\%. The rest of the extra
charge $\approx 20 \%$ is in the $\{ Fe_4\}$ region.}
\label{fig:spheres}
\end{figure}

\begin{figure}[h]
\subfigure[ HOMO 1]{\includegraphics[scale=0.20]{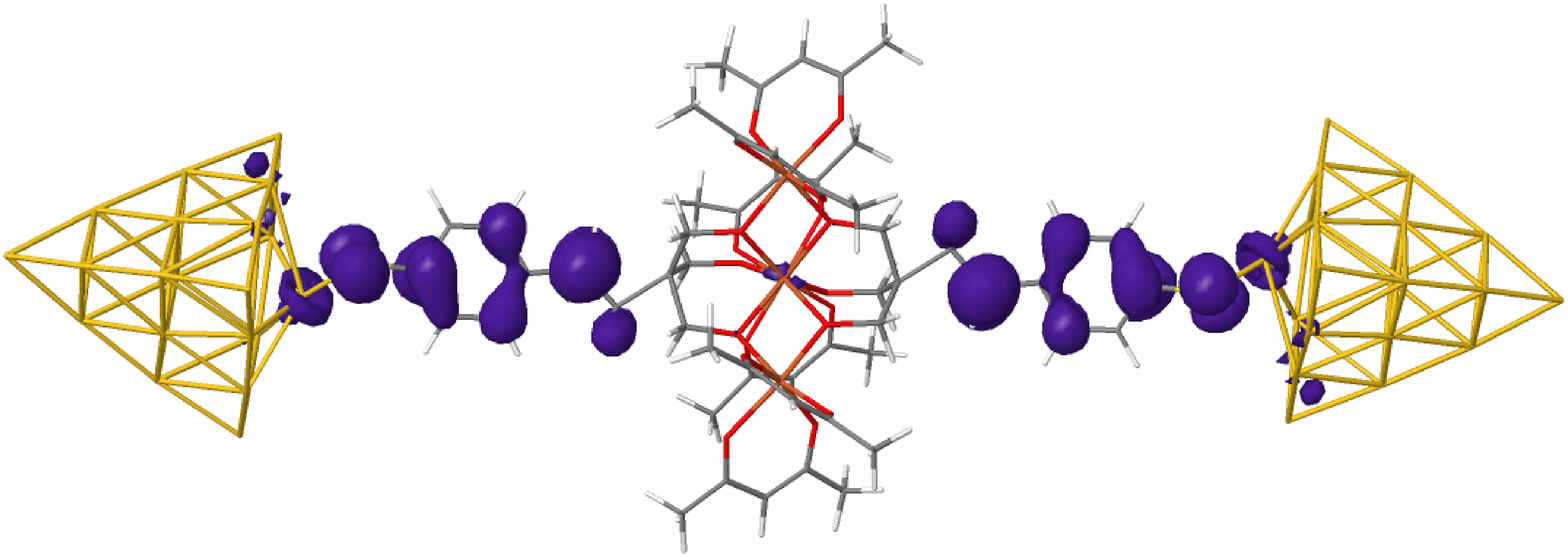}}
\subfigure[ HOMO 2]{\includegraphics[scale=0.20]{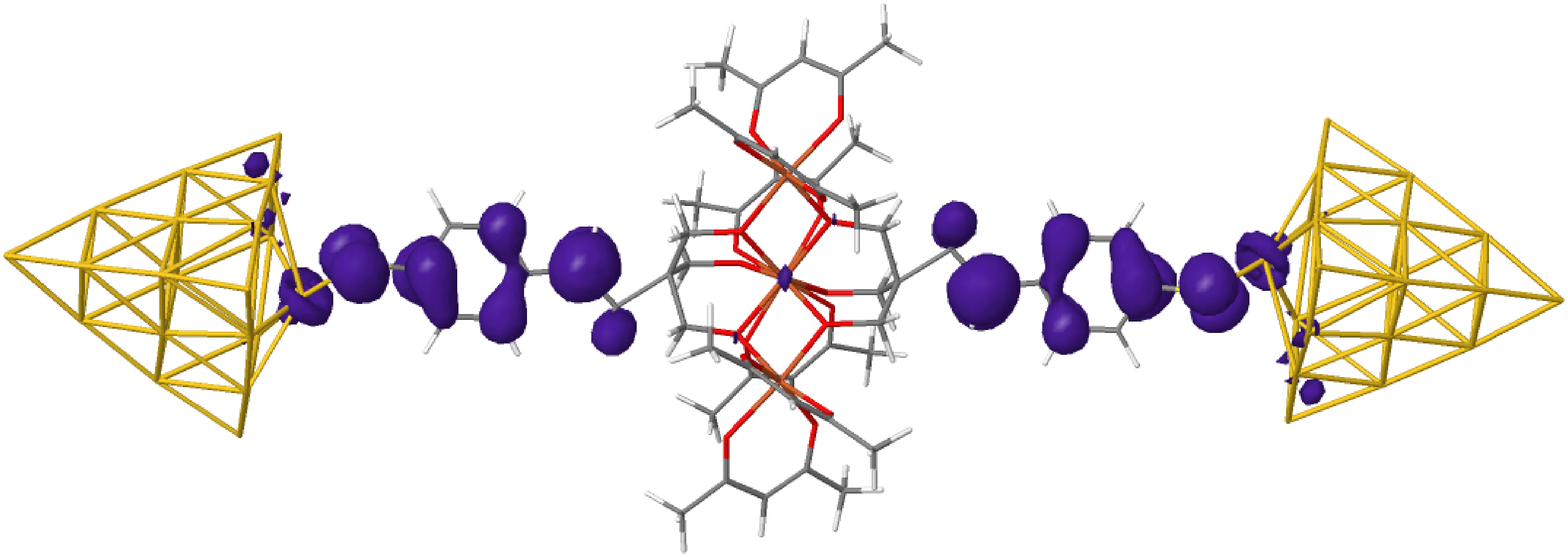}}
\subfigure[ LUMO ]{\includegraphics[scale=0.20]{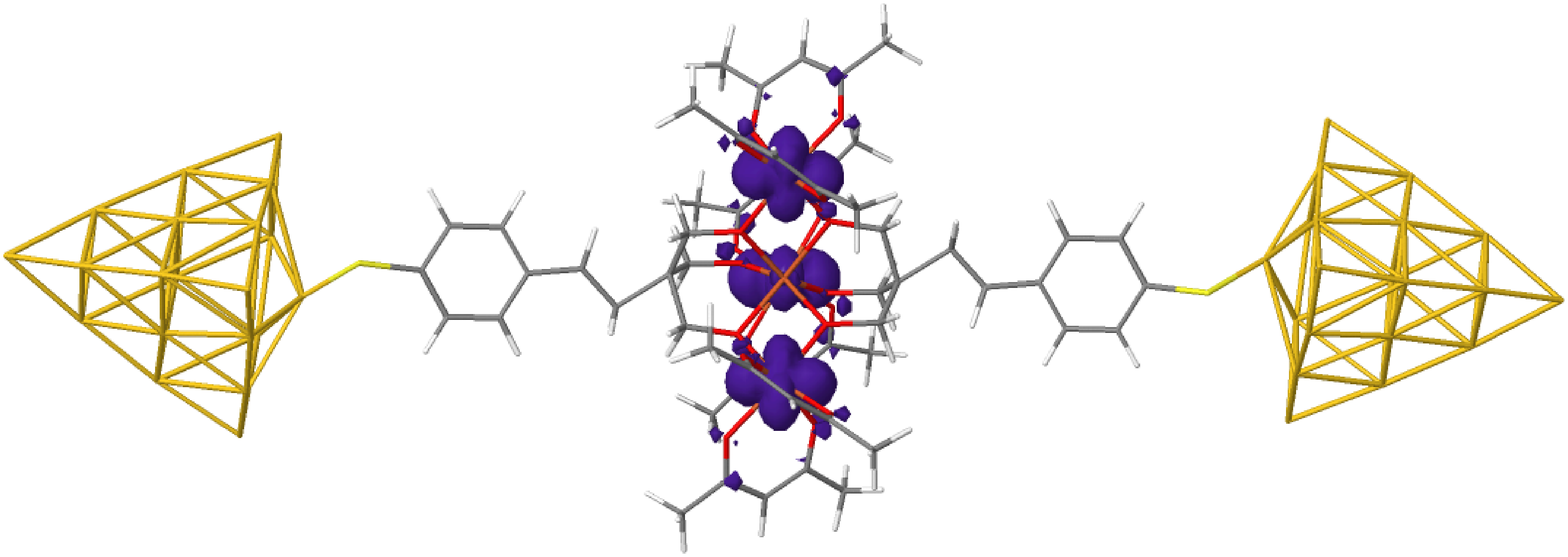}}
\caption{The two quasi-degenerate HOMOs and the LUMO of the anion charge state.
Approximately 20\% of the HOMO wave functions reside on the Au leads, primarily on
the interstitial space between Au atoms. See Fig.~\ref{fig:spheres}.
This contribution is not visible
on the scale of this plot.}
\label{anion_homo1_homo2}
\end{figure}
As shown in the Table~\ref{ch:lead},
in contrast to the case of the isolated $\{Fe_4\}$ SMM where the anisotropy of
the charged states are significantly different from that of the
neutral molecule, when the
leads are attached to the molecule the anisotropy barrier of the charged states
remains close to the value of the
neutral system. We also note
that magnetic properties of the charge states have some dependence
on type of the lead attached to the molecule.

As anticipated above, an explanation of this behavior is already suggested
by the energy diagram of Fig.~\ref{elevel_lead}(b)
and the plots of Fig.~\ref{neutral_homo_lumo}
demonstrating that both the HOMO and the LUMO of the neutral system are states predominately
localized around the Au leads and within the phenyl group respectively.
Therefore, we expect that  when we add or
remove an electron from the system it largely resides in the lead
and phenyl group, leaving the magnetic states in $\{Fe_4\}$ molecule
relatively unchanged. The easy axis, in all cases, points
perpendicularly to the Fe$_4$ plane, as shown in Fig.~\ref{lead}, except
for $Q=-1$ charge state of Type-1 lead, which is in the plane.

Further support to this picture is provided by calculating the real-space location of the extra charge
when an electron is added or subtracted to the system.
As an example, we consider here the case of the anion, where one electron is added to the system.
Since part of this extra charge might end up in interstitial regions between atoms (this is the case
for the extra charge on the Au leads), particular care must be taken in drawing
conclusions based only on the atomic-position plot of the HOMO states,
shown in Fig.~\ref{anion_homo1_homo2}, which might miss this contribution.
To capture the interstitial
contribution, we draw instead a large sphere enclosing a given region of the system.
NRLMOL is able to calculate the extra charge contained globally in that region,
including interstitial contribution.
By repeating the same calculation with different spheres centered at different locations,
we can eventually determine
the amount of extra charge in different relevant parts of the system.

In Fig.~\ref{fig:sA} we consider a sphere (yellow color)
containing  both the lead and the phenyl group linker.
We find that the amount of extra charge contained in this region is 40\% of one electronic charge.
The remaining 20\% is located on atoms in the
nearest surrounding of the $\{Fe_4\}$ core.
In Fig.~\ref{fig:sB} the sphere encloses only the Au lead but no linker. For this case we find
that each Au leads contains 21\% of extra electronic charge. We conclude that when one electron is added
to the system, a total of 42\% of the extra charge resides on the leads,
38\% on the ligands and only 20\% is around the
magnetic core of $\{Fe_4\}$. This 20\% of added charge in not
directly on the Fe atoms and therefore does not change the magnetic properties of $\{Fe_4\}$
significantly.

\begin{figure}[h]
\subfigure[Central Fe atom]{\includegraphics[scale=0.25]{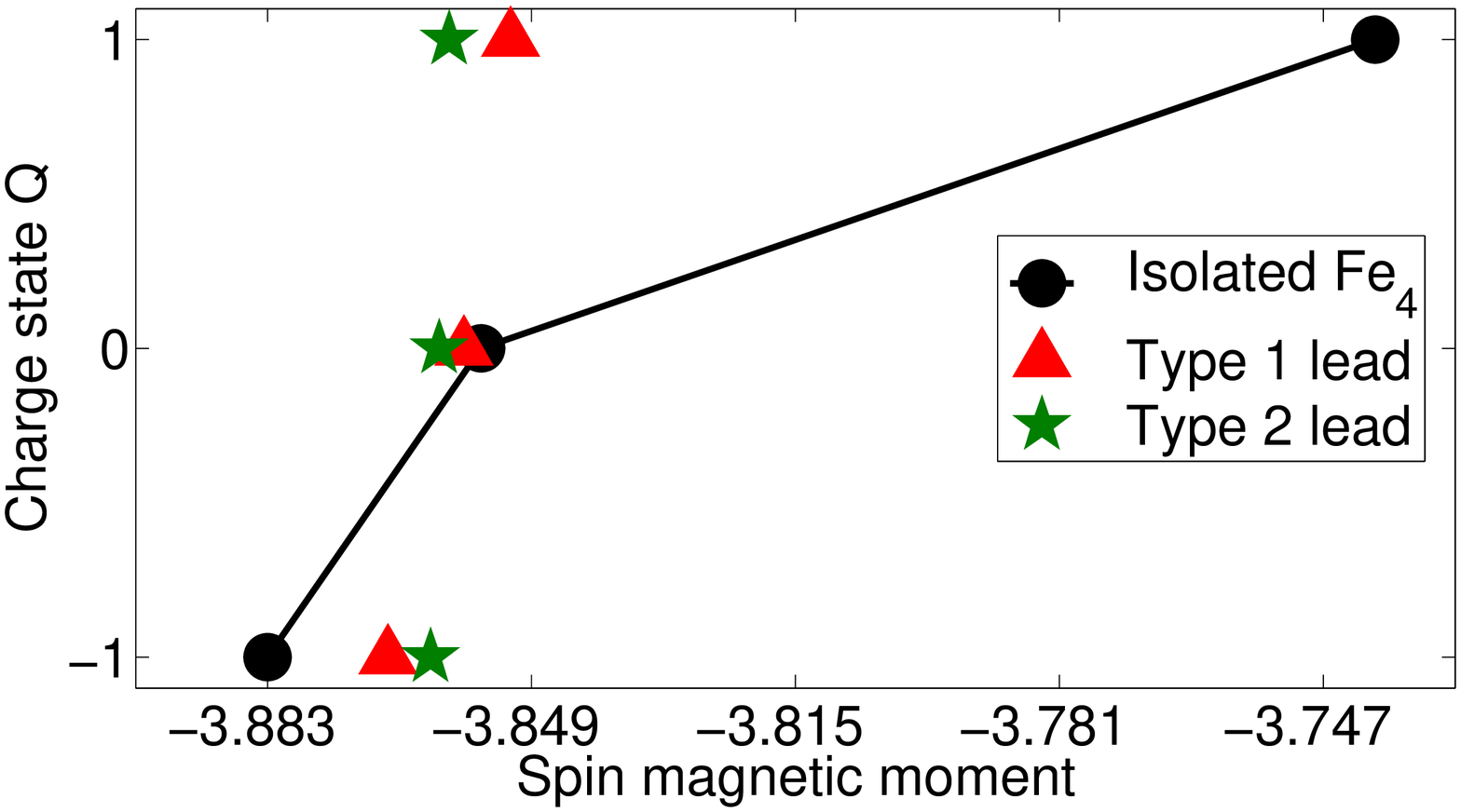}}
\subfigure[Vertex Fe atoms]{\includegraphics[scale=0.25]{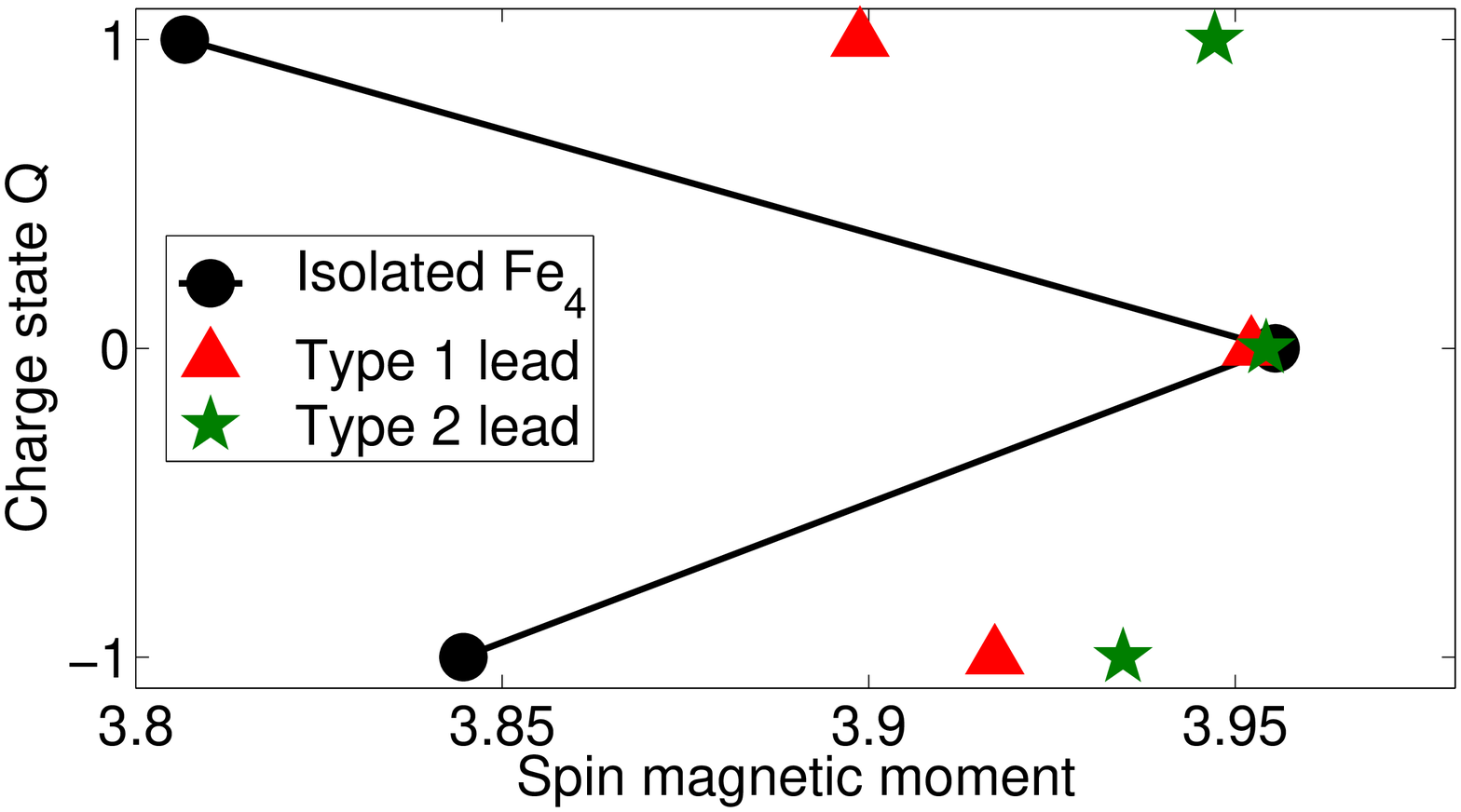}}
\caption{Local magnetic moments of Fe atoms. The negative and the positive
         moments in the two figures imply that moments of central and
         vertex atoms are opposite to each other.}
\label{moments}
\end{figure}

Further evidence of this important conclusion is provided by the comparison of the calculated
local spin magnetic moments of the Fe atoms
for the isolated $\{Fe_4\}$ molecule and for molecule-plus-lead
system, for different charge states. The results are shown in
Fig~\ref{moments}. We note from the figure that for the isolated
molecule the magnetic moments change considerably for the charged
states, whereas for molecule plus lead system, the corresponding change
is very small. Clearly,
when the molecule is attached to the Au leads, adding or removing one unit charge
affects the magnetic states of
the $\{Fe_4\}$ minimally, which is also why we do not see a large change
in magnetic anisotropy for different charge states.

We conclude this section with a few comments on the important issues of the
nature of the charged states and the character of the coupling molecule-leads, as evinced by
the DFT calculations. Firstly, we have seen that when one electron is added or subtracted to
the system $\{Fe_4\}$ + (finite) leads, the extra charge is predominantly localized on the
ligands ($\approx$ 40\%) and on the leads ( $\approx$ 40\%). If we could increase
the size of the leads, a larger fraction of the extra charge would be likely to spread
on the metallic leads.
Therefore one could argue that the charged states $Q= \pm 1$ investigated above
are not a fully adequate description of the charged states involved in the
sequential tunneling processes taking place in a SET, where the additional charge should be
essentially localized on the central island. Secondly, and partly connected to this issue, the non-zero amplitude
of the LUMO wavefunction of the neutral system (which is quite close to the HOMO wavefunction of the anion)
seems to indicate that the ligands considered here do not behave as tunnel barriers of a SET, but rather model
an example of strong coupling between molecule and leads\footnote{Note however that the amplitude of the
LUMO wavefunction is quite small in the central region. Therefore, an electron tunneling in from one of the leads
would still find a bottleneck when tunneling out to the other lead.}
Both these features could be due to limitations of the DFT approach considered here, which tends to over-delocalize
any added charge. Such drawbacks
can possibly be improved by more refined DFT treatments, involving, for example, self-inter corrections.
While we believe that these refinements are important and should be further investigated, the goal and strategy of the
present paper is to simulate with DFT a realistic example of SMM attached to leads, being aware of these limitations.

\section{Effect of an external electric potential}
\label{efield}


In SET devices the charge of the central island weakly coupled to metallic electrodes can be varied
experimentally
one by one
by applying an external electric field via a third gate electrode, which overcomes the charging
energy $e^2/C$ of the island. Here we investigate the effect of an external gate electrode,
whose electric potential tends to confine the extra charge closer to the molecule.
Note that in phenomenological studies of SETs based on model Hamiltonians, a gate voltage only shifts the
energies of the isolated ``quantum dot'' without affecting its wavefunctions.
As we show below, in our case the gate voltage
can be used to localize a wavefunction closer to the molecule and modify its coupling to the leads.
The resulting charged states should be a better representation of the states involved in tunneling
transport in SET when Coulomb blockade is lifted.

We model the external potential by a simple Gaussian confining potential
of the form
\begin{equation}
U= V_0e^{-\alpha_x(x-x_0)^2-\alpha_y(y-y_0)^2-\alpha_z(z-z_0)^2}.
\label{V0}
\end{equation}
Here $V_0$ is magnitude of the potential centered at ($x_0$,$y_0$,$z_0$),
which in our case is the position of the central Fe atom of the $\{Fe_4\}$.
The constants $\alpha$'s are the width of the potential along
the corresponding directions and are chosen so that the potential drops quickly at distances
larger than $\{Fe_4\}$.
The sign of $V_0$ determines whether
the extra electron will be confined into or repelled from the $\{Fe_4\}$
molecule. Thus for the anion case a negative $V_0$ will attract the electron whereas
for the cation case a positive $V_0$ will attract the ``hole'' inside the molecule.

We start by looking
at the effect of the gate voltage on
the anisotropy of the isolated $\{Fe_4\}$ SMM. We have first considered
a gate voltage that depends only on the variable $z$. The resulting
electric field points along the the $z$-axis, which is
the easy axis of the molecule.

\begin{figure}[h]
{\resizebox{3in}{2.0in}{\includegraphics{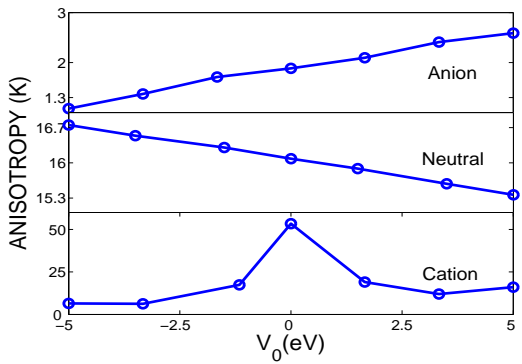}}}
\caption{Variation of the magnetic anisotropy barrier as a function of a confining potential $V_0$
applied along $C_3$  axis (perpendicular to the plane of the $\{Fe_4\}$  triangle) of  an isolated $\{Fe_4\}$ SMM.}
\label{V0_iso}
\end{figure}

From Fig~\ref{V0_iso} we note that in the anion case a confining potential
for the extra electron ($V_0 < 0 $) reduces the anisotropy barrier; whereas repelling the extra
charge away from the molecule increases the anisotropy. The neutral molecule displays an
opposite behavior as a function of $V_0$. In both cases the the behavior of the anisotropy
is close to a linear function of $V_0 < 0 $.
As expected, the variation of the barrier for the neutral molecule is limited, less than 10 \% for the
largest applied voltage. The cation is special. We have seen that at zero voltage, the system has
a large anisotropy barrier
(see Table \ref{ch:iso}).
due to a quasi-degeneracy at the Fermi level.
The external potential lifts this degeneracy and the anisotropy barrier decreases
sharply for both signs
of  $V_0$.

The confining potential that we have applied above
does not break the $C_3$ symmetry of the system and hence the parameter $E$ in Eq.~(\ref{H_D}),
characterizing the transverse
component of the magnetic anisotropy, is
zero. However, the symmetry can be broken by applying an electric field
along directions other than the easy axis. Table~\ref{tb:field}
shows the effect of this broken symmetry on the anisotropy of the
isolated $\{Fe_4\}$ SMM.

\begin{table}[h]
\caption{The effects of confining potential on neutral $\{Fe_4\}$ SMM.
         $\Delta$ is the anisotropy barrier and $D$ and $E$ are the parameters
         of the Hamiltonian Eq.~(\ref{H_D}), all in units of K.}
\label{tb:field}
\begin{tabular}{|c|c|c|c|c|c|c|}   \hline
$V_0$ & $\alpha_x$ & $\alpha_y$ & $\alpha_z$ &   \multicolumn{3}{c|}{ANISOTROPY}  \\  \cline{5-7}
(eV)  &            &          &          &   $\Delta$ (K) & D (K) & E (K)     \\  \hline
0.0   & 0          & 0        & 0        &   16.06        & -0.63 & 0.00      \\  \hline
-5.0  & 0.01       & 0        & 0.01     &   17.00        & -0.64 & -0.03     \\  \hline
-5.0  & 0.01       & 0        & 0        &   15.77        & -0.60 & -0.02     \\  \hline
\end{tabular}
\end{table}

It is evident from Table~\ref{tb:field} that $E$ is no longer zero if
the electric field is applied along directions other than the easy axis.
A non-zero $E$ allows different eigenstates of $z$-component of the giant spin to mix with each
other. a transverse component in principle can cause quantum tunneling of the molecule giant spin.
Thus, this method can be used as electric control of magnetic
properties. It  can play a significant role in transport, for example by modifying spin selections
rule and by opening
alternative channels via quantum tunneling of the magnetization.

We now discuss the effect of the applied gate voltage when the $\{Fe_4\}$  molecule
is attached to Au leads.
In this case we have applied the
field only along the easy axis of the molecule attached to the leads of Type 2,
as shown in Fig.~\ref{lead}. We have seen in the previous
section that since HOMO and LUMO states and states close in energy to these are
primarily localized on the Au leads and phenyl linker, an added or removed
electron leaves $\{Fe_4\}$ largely unaffected. But the presence of a
confining potential ($V_0<$  0 for electrons), applied only on $\{Fe_4\}$ part of the
system, brings the states localized within $\{Fe_4\}$ SMM closer to
LUMO levels. Thus, when an electron is added to the system, the fraction
of this extra charge
that goes inside the molecule increases as we increase the
confining potential. Similarly, when an electron is subtracted from the system, an applied
positive gate
voltage ($V_0>$ 0), tends to localize a fraction of the positive extra charge (a hole)
closer to the molecule.

\begin{figure}[h]
{\resizebox{3in}{1.8in}{\includegraphics{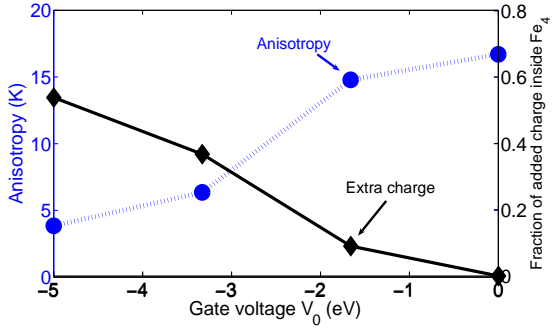}}}
\caption{(Color online) Fraction of the added charge (blue circles) confined inside $\{Fe_4\}$,
         as a function the confining potential strength,
         and corresponding change in the
         magnetic anisotropy (black diamonds) of
         the leads plus molecule system in the anion state, $Q=-1$.
         The external potential
         corresponds to an electric
         field along the $z$-direction. See Fig.~\ref{lead}.}
\label{V0_lead}
\end{figure}

As an example, we consider the effect of a confining potential for
the anion case ($Q = -1$, one electron added to the system).
Fig.~\ref{V0_lead} shows the change in
fractional charge that enters into the $\{Fe_4\}$ molecule as the strength of
confining potential is increased, and the corresponding change in
magnetic anisotropy barrier of the system. Clearly, as the voltage is
increased, more of the added electron is pushed inside the molecule.
As the charge fraction approaches unity, the anisotropy decreases and converges to the value
obtained for the anionic state of the isolated $\{Fe_4\}$ SMM.

It turns out that not only the anisotropy barrier but other magnetic properties converge to
the properties of the anion state of the  isolated $\{Fe_4\}$ as the extra charge, under the effect of
the external potential, moves closer
to the center of the molecule.
The easy axis of the system,
which in zero potential points perpendicular to the plane of the Fe$_4$ triangle and towards the
leads, eventually rotates into the plane of the Fe$_4$ triangle, exactly as in the case of
the anion state of the isolated $\{Fe_4\}$ SMM. Similarly,
as the added charge moves inside the inner magnetic core of $\{Fe_4\}$, the total spin
of the system is reduced from $S= 11/2$ to the value of the anion state of the isolated molecule, $S=9/2$.

Similar results are obtained for the cation. As we apply an increasingly positive voltage, a larger
fraction of a (negative) electron charge
is pushed outside $\{Fe_4\}$, or equivalently, a larger fraction of (positive)
hole is attracted inside
the $\{Fe_4\}$. As a result, the  anisotropy barrier increases and it reaches a value of 22.8 K for
$V_0 = 5 $ eV, with more than half of the extra (positive) charge now inside $\{Fe_4\}$.
Similarly, the spin also switches from $S=11/2$ at $V_0 = 0 $ to $S=9/2$ at $V_0 = 5 $ eV.
Again, this is consistent with both the spin and the anisotropy converging
towards the corresponding values of the isolated cation state.

\begin{figure}[h]
{\resizebox{3in}{1.8in}{\includegraphics{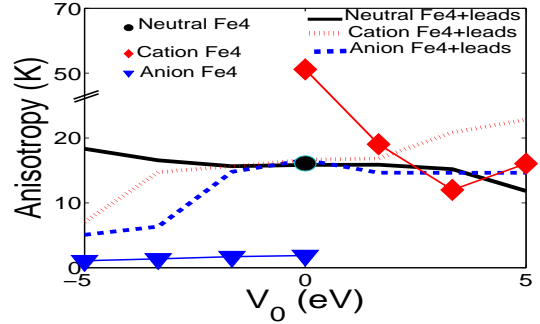}}}
\caption{ Magnetic anisotropy as a function of a confining
         field for the molecule plus leads (Type 2) system, for the three charge
         states. The field is applied
          along the easy axis, that is, perpendicular to the plane of the molecule.
          The anisotropy for the the three charge states for the isolated molecule
           is included as a comparison (see diamond, circle and triangle symbols).}
\label{V0_lead_all_comp2}
\end{figure}

A summary of the dependence of the anisotropy barrier as a function of the external potential
for all three charge state in shown in
Fig.~\ref{V0_lead_all_comp2}. While the anisotropy of neutral state displays
a weak dependence on the field,
the anisotropy of the two charged states is significantly affected.
These calculations demonstrate that, for a SET with a  $\{Fe_4\}$ SMM as a central island,
by manipulating the position of the additional charge with a gate voltage, it is possible to modify
the magnetic properties of the SMM. This in turn can have important effects on the tunneling
transport of the device.


\section{Comparison with SET experiments}
\label{expe}


Recent SET experiments\cite{alexander10,vzant12} have permitted the first measurements
of the magnetic characteristics of the $\{Fe_4\}$ SMM weakly coupled to Au leads, for both neutral
and charged states.  Comparison with our theoretical results has be done with caution, since
important details (e.g., the type of linker used, see below) might differ in the two cases.
In a first study  Zyazin {\it et al.}\cite{alexander10} have studied quantum transport in the
inelastic cotunneling regime. By measuring the zero-field splitting of magnetic excitations
and
their dependence on the magnetic field, it was possible, with help of the model Hamiltonian
of Eq.~(\ref{H_D}), to extract values of the giant spin $S$ and the magnetic anisotropy
barrier $\Delta = D S^2$ for three adjacent charge states, $N$ (neutral), $N+1$ (anion) and
$N-1$ (cation). For the neutral state, the spin of the molecule was found to be equal to expected
value $S_N=5$ and the anisotropy barrier to be consistent with the value in the bulk phase,
$\Delta_{N} = 1.4 {\rm meV} = 16.24$K.
These results are in good agreement with our theoretical estimate
for the $Q=0$ charge state, see Table~\ref{ch:lead}.

For the charged states, the situation is more complicated.
For the reduced molecule (anion state, one electron added)
the experimental measurements gave $S_{N+1} = 11/2$ for the spin
and $\Delta_{N+1} = 2.7 {\rm meV} = 31.30$K
for the anisotropy. For the oxidized molecule (cation state, one electron removed),
the measurements gave $S_{N-1} = 9/2$ and $\Delta_{N-1} = 1.8 {\rm meV} = 20.90\ K$ respectively.
Comparing these findings with the results of our calculations
(see Table~\ref{ch:lead}))
we can see that,
for a given choice of lead type (Type 1 or Type 2),
the theoretical value of
the spin is consistent with the experimental one for either cation or anion, but not for both.
We can also conclude that the experiment typically finds
larger values for the anisotropy barrier, for both reduced and oxidized states, than
the values predicted by theory.

Several reasons can be responsible for these discrepancies.
First, the functionalization  of the molecule used in the experiment is slightly different from the
one used in the calculations. In fact, in the experiments two different types of ligands were used.
In one case (labeled as sample A) the $\{Fe_4\}$ SMM was connected to the Au lead
the phenyl group. In the second case, (labeled as sample B) the $\{Fe_4\}$ was
connected via a thiol group, C$_9$S. The coupling molecule-lead turns out to be weaker in
sample A than in sample B.\footnote{Note that in Ref.~\onlinecite{alexander10},
the results reported for the anion and cation
were obtained for sample A and B respectively.} As shown above, in all our calculations
we have used a different type of linker, which was
combination of a phenyl group and a thiol group.

Secondly, in experiment different charge states are achieved by
adding or removing electrons to the central region of SET via a gate
voltage. In the theoretical calculations, the two relevant charged
states are constructing by adding or removing an electron to a
system consisting of the $\{Fe_4\}$ connected to finite leads. The
extra charge is allowed to relax in the self-consistent field, and
it occupies regions away  from the $\{Fe_4\}$, which affects the
magnetic properties of the system. Indeed, confining on the SMM with
an external gate modifies the anisotropy barrier.

Third, the evaluation of the magnetic properties from experiment
done in Ref.~\onlinecite{alexander10} relies on the use of the model Hamiltonian of Eq.~(\ref{H_D}).
The fitting of the experimental
results maintains a degree of uncertainty and arbitrariness,
and moreover it could be problematic in cases of level degeneracy at the
Fermi level,
not uncommon for charged states. In this case, we have seen the the giant-spin model
of Eq.~(\ref{H_D}) might become inadequate.

Finally, the method of Ref.~\onlinecite{alexander10} relies on the
measurement of inelastic cotunneling excitations, which is quite
sensitive to the coupling between molecule and leads, and therefore
it is a procedure not immune of uncertainties. Indeed, in a more
recent paper Burzur{\'\i} {\it et al.}\cite{vzant12} introduced a
novel gate-voltage spectroscopy technique which permits the
measurement of the anisotropy of an individual SMM in different
charge states by tracking the dependence of the charge degeneracy
point as a function of magnetic field. This method is much more
sensitive and accurate than the method based on conventional
transport spectroscopy employed in Ref.~\onlinecite{alexander10}.
The spin Hamiltonian provides a good fit of the data if $S_N =5$,
$\Delta_{N}= 16.2$K for the neutral state, $S_{N+1}=9/2 $ and
$\Delta_{N+1}= 16.0$K for the reduced state and $S_{N-1}=11/2$ and
$\Delta_{N-1} = 16.5$K for the oxides state. Furthermore, the
orientation of the easy axis is found to exhibit only small
variations among different charge states. Although some details
might differ, \footnote{In general, in Ref.~\onlinecite{vzant12} it
is found that upon reduction, either from $N \rightarrow N+1$ or
from $N-1 \rightarrow N$, the spin $S$ always {\it decreases} and
the anisotropy parameter $D$, defined via $\Delta = D S^2$, always
{\it increases}. The results of our calculations (see
Table~\ref{ch:lead}) show that both the anion and the cation have
preferably $S= 11/2$ for a lead of Type-2, whereas $S= 9/2$ for a
lead of Type-1. However, states with swapped spin configurations $S=
11/2 \leftrightarrow S= 9/2$ are quite close in energy for both
charged states.}, these results are quite consistent with the small
variations in anisotropy magnitude and the unchanged orientation of
the easy axis that we find for the three charge states in our
theoretical analysis. See Table~\ref{ch:lead}.

As we mentioned above, the small variation in the anisotropy for
different charged states found in our calculations is related to the
fact that any extra charge added to the molecule + leads system
tends not to reside directly on the magnetic atoms, but mainly on
the ligands and on the Au leads. We pointed out that this could be,
in part, an artificial effect due to the way we constructed charged
states in our finite-size system and to the delocalizing character
of our DFT approach. On the other hand, it is interesting that our
estimates for the magnetic anisotropy are essentially consistent
with the experimental results of Ref.~\onlinecite{vzant12}, which
are obtained exactly at charge degeneracy points. At these special
values of the external gate voltage the energy of two adjacent
charge states is the same. We can imagine that at the degeneracy
point the extra charge can swap energy-free from the electrodes and
the molecule and might be localized primarily in region where the
molecule is connected to the leads. If this were the case, the
charge distribution shown in Fig.~\ref{fig:spheres} could be in fact
a more realistic description of the charge states than previously
anticipated. We could also surmise that, exactly as it happens in
our calculations when we apply an external confining potential,
changing the gate voltage to move away from the degeneracy point and
make a given charge state more stable could strongly affect the the
magnetic anisotropy. Indeed the results of the the first
experiment\cite{alexander10}, where the anisotropy was not extracted
at degeneracy points but in the middle of a Coulomb blockade
diamond, show a significantly enhanced anisotropy for charged
states. The two experimental results could simply indicate that in
one case the extra charge is localized closer to the SMM than in the
other, exactly as it happens in our theoretical modeling.

\section{Conclusions}
\label{conclusion}



In this  paper we have studied the electronic and magnetic properties of a
$\{Fe_4\}$ SMM in a single-electron transistor (SET) geometry,
using DFT as implemented in NRLMOL.
We have modeled the system by a $\{Fe_4\}$ functionalized
with phenyl groups attached to two metals leads described by
Au$_{20}$ nanoclusters. Our calculations show that
the magnetic structure of the neutral $\{Fe_4\}$ SMM, that is, its
spin ordering and magnetic anisotropy, remains stable in the
presence of metallic leads. Specifically the ground state spin is $S=5$ and
the anisotropy barrier is of the order of $16 K$,
like for the isolated $\{Fe_4\}$.
This result is ascribed to the fact
that, when attaching leads to $\{Fe_4\}$, any charge transfer between
the molecule and the metal leads occurs primarily in the contact region and on
the ligands, but does
involve the magnetic core of the molecule.

Based on the properties of the HOMO and LUMO of the neutral system,
when an electron is added or subtracted to the molecule-lead system,
we find that the added charge ($Q= \mp 1 $) is primarily located on the ligands and on the
leads. As a result, while the total spin of this finite system changes
by $\Delta S = \pm 1/2$, the magnetic anisotropy displays small variations
both in magnitude and orientation with respect to the neutral state.
In contrast, the anisotropy of the anion and cation states
of an isolated $\{Fe_4\}$ is quite different from the values of neutral
molecule, since the added extra charge penetrates the region of the Fe atoms.
The theoretical study of charged states $Q = \pm 1$ for the molecule-leads system is technically
challenging, due to occurrence of small HOMO-LUMO gaps and consequent fractional
occupancies of the states around the Fermi level. Furthermore, DFT tends to over-delocalize
any added charge in the peripheral parts of the system. Nevertheless, the analysis of these
states presented here sheds light on the properties of a $\{Fe_4\}$ SET when individual electrons are
added or subtracted to the ``quantum dot'' by overcoming the Coulomb charging energy with a gate voltage.

We have shown that an external electric potential, modeling a gate voltage,
can be used to manipulate the charge on the molecule-leads system and with that
the magnetic properties of the device. In particular, for the two charged states $Q = \pm 1$
when the extra charge, under the effect of the potential, is progressively removed from the ligands-leads region into the magnetic
core of the molecule, the magnetic properties converge to the properties of the anion and cation states of the isolated $\{Fe_4\}$.
This is an example of the electric control of magnetism of a SMM in a SET.
The charged states of the molecule-leads system
in the presence of external fields studied  in this paper can be used to construct the transition matrix
elements entering a quantum master equation
describing tunneling transport in a SET. With the limitations inherent to the DFT approach mentioned above, these
states incorporate charging effects for the SMM weakly coupled to metal leads.

We have compared the results of our numerical calculations with the results
of two recent experimental studies of tunneling transport in a $\{Fe_4\}$ three-terminal device in the
Coulomb blockade regime\cite{alexander10,vzant12}. This comparison must be made with caution since some important details
(e.g., the precise atomic and electronic structure of the ligands) are different and might
explain some of the discrepancies between theory and experiment that we find. Nevertheless, one of these experiments\cite{vzant12}
finds that the anisotropy for the two charged states $Q =\pm 1$ displays only small variations in magnitude and orientation from the
corresponding values of the neutral states, in agreement with our theoretical findings. Interestingly enough, the experimental
values are extracted by tracking the dependence of the charge degeneracy point between two adjacent charged states as a function
of the magnetic field. Our numerical calculations show that the nearly-independence of the magnetic anisotropy on the charged states
is related to the position of the added charge being far away from the magnetic core of the molecule.
Thus the agreement between theory ad experiment might indicate that for a $\{Fe_4\}$ SET the charge of a
added (subtracted) electron close to a
charge-degeneracy point is primarily located on the ligands and in the contact region with the leads. If correct, this would be
an example in which a magnetic property of SMM-based SET can provide information on the electronic properties of the charged states.

For molecule-lead systems with finite gaps, we expect our results to provide
accurate predictions of experiment. However for those cases where
HOMO-LUMO gaps are very small and the electronic states at the Fermi level are partially
occupied, further understanding will require variationally accounting
for the electronic occupations along the lines suggested in Ref.~\cite{pedjak91}
Another point of view is that such fractionally occupied solutions are also
strongly affected by self-interaction corrections and that accounting for such
corrections will often significantly decrease the possibility of fractionally
occupied solutions. Self-interaction corrections are also likely to provide a more complete
understanding of the nature of charged states investigate in this paper. Addressing spin-dependent conductance in macro-molecular
to meso-scale devices will require efficient solutions to these problems and
renewed efforts at extracting quantitative model Hamiltonians for such systems.

\section*{Acknowledgments}
This work was supported by the faculty of Natural Science and Technology
at Linnaeus University,
the Swedish Research Council under Grants No: 621-2007-5019 and 621-2010-3761,
and the NordForsk research network 080134 ``Nanospintronics: theory and simulations".

\bibliography{Fe4JN}

\begin{thebibliography}{42}%
\makeatletter
\providecommand \@ifxundefined [1]{%
 \@ifx{#1\undefined}
}%
\providecommand \@ifnum [1]{%
 \ifnum #1\expandafter \@firstoftwo
 \else \expandafter \@secondoftwo
 \fi
}%
\providecommand \@ifx [1]{%
 \ifx #1\expandafter \@firstoftwo
 \else \expandafter \@secondoftwo
 \fi
}%
\providecommand \natexlab [1]{#1}%
\providecommand \enquote  [1]{``#1''}%
\providecommand \bibnamefont  [1]{#1}%
\providecommand \bibfnamefont [1]{#1}%
\providecommand \citenamefont [1]{#1}%
\providecommand \href@noop [0]{\@secondoftwo}%
\providecommand \href [0]{\begingroup \@sanitize@url \@href}%
\providecommand \@href[1]{\@@startlink{#1}\@@href}%
\providecommand \@@href[1]{\endgroup#1\@@endlink}%
\providecommand \@sanitize@url [0]{\catcode `\\12\catcode `\$12\catcode
  `\&12\catcode `\#12\catcode `\^12\catcode `\_12\catcode `\%12\relax}%
\providecommand \@@startlink[1]{}%
\providecommand \@@endlink[0]{}%
\providecommand \url  [0]{\begingroup\@sanitize@url \@url }%
\providecommand \@url [1]{\endgroup\@href {#1}{\urlprefix }}%
\providecommand \urlprefix  [0]{URL }%
\providecommand \Eprint [0]{\href }%
\providecommand \doibase [0]{http://dx.doi.org/}%
\providecommand \selectlanguage [0]{\@gobble}%
\providecommand \bibinfo  [0]{\@secondoftwo}%
\providecommand \bibfield  [0]{\@secondoftwo}%
\providecommand \translation [1]{[#1]}%
\providecommand \BibitemOpen [0]{}%
\providecommand \bibitemStop [0]{}%
\providecommand \bibitemNoStop [0]{.\EOS\space}%
\providecommand \EOS [0]{\spacefactor3000\relax}%
\providecommand \BibitemShut  [1]{\csname bibitem#1\endcsname}%
\let\auto@bib@innerbib\@empty
\bibitem [{\citenamefont {Rocha}\ \emph {et~al.}(2005)\citenamefont {Rocha},
  \citenamefont {Garcia-Suarez}, \citenamefont {Bailey}, \citenamefont
  {Lambert}, \citenamefont {Ferrer},\ and\ \citenamefont {Sanvito}}]{rocha05}%
  \BibitemOpen
  \bibfield  {author} {\bibinfo {author} {\bibfnamefont {A.~R.}\ \bibnamefont
  {Rocha}}, \bibinfo {author} {\bibfnamefont {V.~M.}\ \bibnamefont
  {Garcia-Suarez}}, \bibinfo {author} {\bibfnamefont {S.~W.}\ \bibnamefont
  {Bailey}}, \bibinfo {author} {\bibfnamefont {C.~J.}\ \bibnamefont {Lambert}},
  \bibinfo {author} {\bibfnamefont {J.}~\bibnamefont {Ferrer}}, \ and\ \bibinfo
  {author} {\bibfnamefont {S.}~\bibnamefont {Sanvito}},\ }\href@noop {}
  {\bibfield  {journal} {\bibinfo  {journal} {Nat. Mater.}\ }\textbf {\bibinfo
  {volume} {4}},\ \bibinfo {pages} {335} (\bibinfo {year} {2005})}\BibitemShut
  {NoStop}%
\bibitem [{\citenamefont {Rocha}\ \emph {et~al.}(2006)\citenamefont {Rocha},
  \citenamefont {Garcia-Suarez}, \citenamefont {Bailey}, \citenamefont
  {Lambert}, \citenamefont {Ferrer},\ and\ \citenamefont {Sanvito}}]{rocha06}%
  \BibitemOpen
  \bibfield  {author} {\bibinfo {author} {\bibfnamefont {A.~R.}\ \bibnamefont
  {Rocha}}, \bibinfo {author} {\bibfnamefont {V.~M.}\ \bibnamefont
  {Garcia-Suarez}}, \bibinfo {author} {\bibfnamefont {S.}~\bibnamefont
  {Bailey}}, \bibinfo {author} {\bibfnamefont {C.}~\bibnamefont {Lambert}},
  \bibinfo {author} {\bibfnamefont {J.}~\bibnamefont {Ferrer}}, \ and\ \bibinfo
  {author} {\bibfnamefont {S.}~\bibnamefont {Sanvito}},\ }\href@noop {}
  {\bibfield  {journal} {\bibinfo  {journal} {Phys. Rev. B}\ }\textbf {\bibinfo
  {volume} {73}},\ \bibinfo {pages} {085414} (\bibinfo {year}
  {2006})}\BibitemShut {NoStop}%
\bibitem [{\citenamefont {Trif}\ \emph {et~al.}(2010)\citenamefont {Trif},
  \citenamefont {Troiani}, \citenamefont {Stepanenko},\ and\ \citenamefont
  {Loss}}]{trif10}%
  \BibitemOpen
  \bibfield  {author} {\bibinfo {author} {\bibfnamefont {M.}~\bibnamefont
  {Trif}}, \bibinfo {author} {\bibfnamefont {F.}~\bibnamefont {Troiani}},
  \bibinfo {author} {\bibfnamefont {D.}~\bibnamefont {Stepanenko}}, \ and\
  \bibinfo {author} {\bibfnamefont {D.}~\bibnamefont {Loss}},\ }\href@noop {}
  {\bibfield  {journal} {\bibinfo  {journal} {Phys. Rev. B}\ }\textbf {\bibinfo
  {volume} {82}},\ \bibinfo {pages} {045429} (\bibinfo {year}
  {2010})}\BibitemShut {NoStop}%
\bibitem [{\citenamefont {Wernsdorfer}\ and\ \citenamefont
  {Sessoli}(1999)}]{sessoli99}%
  \BibitemOpen
  \bibfield  {author} {\bibinfo {author} {\bibfnamefont {W.}~\bibnamefont
  {Wernsdorfer}}\ and\ \bibinfo {author} {\bibfnamefont {R.}~\bibnamefont
  {Sessoli}},\ }\href@noop {} {\bibfield  {journal} {\bibinfo  {journal}
  {Science}\ }\textbf {\bibinfo {volume} {284}},\ \bibinfo {pages} {133}
  (\bibinfo {year} {1999})}\BibitemShut {NoStop}%
\bibitem [{\citenamefont {Bogani}\ and\ \citenamefont
  {Wernsdorfer}(2008)}]{bogani08}%
  \BibitemOpen
  \bibfield  {author} {\bibinfo {author} {\bibfnamefont {L.}~\bibnamefont
  {Bogani}}\ and\ \bibinfo {author} {\bibfnamefont {W.}~\bibnamefont
  {Wernsdorfer}},\ }\href@noop {} {\bibfield  {journal} {\bibinfo  {journal}
  {Nat. Mater.}\ }\textbf {\bibinfo {volume} {7}},\ \bibinfo {pages} {179}
  (\bibinfo {year} {2008})}\BibitemShut {NoStop}%
\bibitem [{\citenamefont {Candini}\ \emph {et~al.}(2011)\citenamefont
  {Candini}, \citenamefont {Klyatskaya}, \citenamefont {Ruben}, \citenamefont
  {Wernsdorfer},\ and\ \citenamefont {Affronte}}]{candini11}%
  \BibitemOpen
  \bibfield  {author} {\bibinfo {author} {\bibfnamefont {A.}~\bibnamefont
  {Candini}}, \bibinfo {author} {\bibfnamefont {S.}~\bibnamefont {Klyatskaya}},
  \bibinfo {author} {\bibfnamefont {M.}~\bibnamefont {Ruben}}, \bibinfo
  {author} {\bibfnamefont {W.}~\bibnamefont {Wernsdorfer}}, \ and\ \bibinfo
  {author} {\bibfnamefont {M.}~\bibnamefont {Affronte}},\ }\href@noop {}
  {\bibfield  {journal} {\bibinfo  {journal} {Nano Letters}\ }\textbf {\bibinfo
  {volume} {11}},\ \bibinfo {pages} {2634} (\bibinfo {year}
  {2011})}\BibitemShut {NoStop}%
\bibitem [{\citenamefont {Gatteschi}\ \emph {et~al.}(2006)\citenamefont
  {Gatteschi}, \citenamefont {Sessoli},\ and\ \citenamefont
  {Villain}}]{sessoli06}%
  \BibitemOpen
  \bibfield  {author} {\bibinfo {author} {\bibfnamefont {D.}~\bibnamefont
  {Gatteschi}}, \bibinfo {author} {\bibfnamefont {R.}~\bibnamefont {Sessoli}},
  \ and\ \bibinfo {author} {\bibfnamefont {J.}~\bibnamefont {Villain}},\
  }\href@noop {} {\emph {\bibinfo {title} {Molecular Nenomagnets}}}\ (\bibinfo
  {publisher} {Oxford University Press, Oxford},\ \bibinfo {year}
  {2006})\BibitemShut {NoStop}%
\bibitem [{\citenamefont {Heersche}\ \emph {et~al.}(2006)\citenamefont
  {Heersche}, \citenamefont {de~Groot}, \citenamefont {Folk}, \citenamefont
  {van~der Zant}, \citenamefont {Romeike}, \citenamefont {Wegewijs},
  \citenamefont {Zobbi}, \citenamefont {Barreca}, \citenamefont {Tondello},\
  and\ \citenamefont {Cornia}}]{heersche06}%
  \BibitemOpen
  \bibfield  {author} {\bibinfo {author} {\bibfnamefont {H.~B.}\ \bibnamefont
  {Heersche}}, \bibinfo {author} {\bibfnamefont {Z.}~\bibnamefont {de~Groot}},
  \bibinfo {author} {\bibfnamefont {J.~A.}\ \bibnamefont {Folk}}, \bibinfo
  {author} {\bibfnamefont {H.~S.~J.}\ \bibnamefont {van~der Zant}}, \bibinfo
  {author} {\bibfnamefont {C.}~\bibnamefont {Romeike}}, \bibinfo {author}
  {\bibfnamefont {M.~R.}\ \bibnamefont {Wegewijs}}, \bibinfo {author}
  {\bibfnamefont {L.}~\bibnamefont {Zobbi}}, \bibinfo {author} {\bibfnamefont
  {D.}~\bibnamefont {Barreca}}, \bibinfo {author} {\bibfnamefont
  {E.}~\bibnamefont {Tondello}}, \ and\ \bibinfo {author} {\bibfnamefont
  {A.}~\bibnamefont {Cornia}},\ }\href@noop {} {\bibfield  {journal} {\bibinfo
  {journal} {Phys. Rev. Lett.}\ }\textbf {\bibinfo {volume} {96}},\ \bibinfo
  {pages} {206801} (\bibinfo {year} {2006})}\BibitemShut {NoStop}%
\bibitem [{\citenamefont {Jo}\ \emph {et~al.}(2006)\citenamefont {Jo},
  \citenamefont {Grose}, \citenamefont {Baheti}, \citenamefont {Deshmukh},
  \citenamefont {Sokol}, \citenamefont {Rumberger}, \citenamefont
  {Hendrickson}, \citenamefont {Long}, \citenamefont {Park},\ and\
  \citenamefont {Ralph}}]{jo06}%
  \BibitemOpen
  \bibfield  {author} {\bibinfo {author} {\bibfnamefont {M.-H.}\ \bibnamefont
  {Jo}}, \bibinfo {author} {\bibfnamefont {J.~E.}\ \bibnamefont {Grose}},
  \bibinfo {author} {\bibfnamefont {K.}~\bibnamefont {Baheti}}, \bibinfo
  {author} {\bibfnamefont {M.~M.}\ \bibnamefont {Deshmukh}}, \bibinfo {author}
  {\bibfnamefont {J.~J.}\ \bibnamefont {Sokol}}, \bibinfo {author}
  {\bibfnamefont {E.~M.}\ \bibnamefont {Rumberger}}, \bibinfo {author}
  {\bibfnamefont {D.~N.}\ \bibnamefont {Hendrickson}}, \bibinfo {author}
  {\bibfnamefont {J.~R.}\ \bibnamefont {Long}}, \bibinfo {author}
  {\bibfnamefont {H.}~\bibnamefont {Park}}, \ and\ \bibinfo {author}
  {\bibfnamefont {D.~C.}\ \bibnamefont {Ralph}},\ }\href@noop {} {\bibfield
  {journal} {\bibinfo  {journal} {Nano Lett.}\ }\textbf {\bibinfo {volume}
  {6}},\ \bibinfo {pages} {2014} (\bibinfo {year} {2006})}\BibitemShut
  {NoStop}%
\bibitem [{\citenamefont {Sessoli}\ \emph {et~al.}(1993)\citenamefont
  {Sessoli}, \citenamefont {Tsai}, \citenamefont {Schake}, \citenamefont
  {Wang}, \citenamefont {Vincent}, \citenamefont {Folting}, \citenamefont
  {Gatteschi}, \citenamefont {Christou},\ and\ \citenamefont
  {Hendrickson}}]{sessoli93}%
  \BibitemOpen
  \bibfield  {author} {\bibinfo {author} {\bibfnamefont {R.}~\bibnamefont
  {Sessoli}}, \bibinfo {author} {\bibfnamefont {H.}~\bibnamefont {Tsai}},
  \bibinfo {author} {\bibfnamefont {A.}~\bibnamefont {Schake}}, \bibinfo
  {author} {\bibfnamefont {S.~Y.}\ \bibnamefont {Wang}}, \bibinfo {author}
  {\bibfnamefont {J.~B.}\ \bibnamefont {Vincent}}, \bibinfo {author}
  {\bibfnamefont {K.}~\bibnamefont {Folting}}, \bibinfo {author} {\bibfnamefont
  {D.}~\bibnamefont {Gatteschi}}, \bibinfo {author} {\bibfnamefont
  {G.}~\bibnamefont {Christou}}, \ and\ \bibinfo {author} {\bibfnamefont
  {D.~N.}\ \bibnamefont {Hendrickson}},\ }\href@noop {} {\bibfield  {journal}
  {\bibinfo  {journal} {J. Am. Chem. Soc.}\ }\textbf {\bibinfo {volume}
  {115}},\ \bibinfo {pages} {1804} (\bibinfo {year} {1993})}\BibitemShut
  {NoStop}%
\bibitem [{\citenamefont {Mannini}\ \emph {et~al.}(2008)\citenamefont
  {Mannini}, \citenamefont {Sainctavit}, \citenamefont {Sessoli}, \citenamefont
  {Cartier~dit Moulin}, \citenamefont {Pineider}, \citenamefont {Arrio},
  \citenamefont {Cornia},\ and\ \citenamefont {Gatteschi}}]{sessoli08}%
  \BibitemOpen
  \bibfield  {author} {\bibinfo {author} {\bibfnamefont {M.}~\bibnamefont
  {Mannini}}, \bibinfo {author} {\bibfnamefont {P.}~\bibnamefont {Sainctavit}},
  \bibinfo {author} {\bibfnamefont {R.}~\bibnamefont {Sessoli}}, \bibinfo
  {author} {\bibfnamefont {C.}~\bibnamefont {Cartier~dit Moulin}}, \bibinfo
  {author} {\bibfnamefont {F.}~\bibnamefont {Pineider}}, \bibinfo {author}
  {\bibfnamefont {M.-A.}\ \bibnamefont {Arrio}}, \bibinfo {author}
  {\bibfnamefont {A.}~\bibnamefont {Cornia}}, \ and\ \bibinfo {author}
  {\bibfnamefont {D.}~\bibnamefont {Gatteschi}},\ }\href@noop {} {\bibfield
  {journal} {\bibinfo  {journal} {Chemistry – A European Journal}\ }\textbf
  {\bibinfo {volume} {14}},\ \bibinfo {pages} {7530} (\bibinfo {year}
  {2008})}\BibitemShut {NoStop}%
\bibitem [{\citenamefont {Accorsi}\ \emph
  {et~al.}(2006{\natexlab{a}})\citenamefont {Accorsi}, \citenamefont {Barra},
  \citenamefont {Caneschi}, \citenamefont {Chastanet}, \citenamefont {Cornia},
  \citenamefont {Fabretti}, \citenamefont {Gatteschi}, \citenamefont {Mortalo},
  \citenamefont {Olivieri}, \citenamefont {Parenti}, \citenamefont {Rosa},
  \citenamefont {Sessoli}, \citenamefont {Sorace}, \citenamefont
  {Wernsdorfer},\ and\ \citenamefont {Zobbi}}]{accorsi_2006}%
  \BibitemOpen
  \bibfield  {author} {\bibinfo {author} {\bibfnamefont {S.}~\bibnamefont
  {Accorsi}}, \bibinfo {author} {\bibfnamefont {A.-L.}\ \bibnamefont {Barra}},
  \bibinfo {author} {\bibfnamefont {A.}~\bibnamefont {Caneschi}}, \bibinfo
  {author} {\bibfnamefont {G.}~\bibnamefont {Chastanet}}, \bibinfo {author}
  {\bibfnamefont {A.}~\bibnamefont {Cornia}}, \bibinfo {author} {\bibfnamefont
  {A.~C.}\ \bibnamefont {Fabretti}}, \bibinfo {author} {\bibfnamefont
  {D.}~\bibnamefont {Gatteschi}}, \bibinfo {author} {\bibfnamefont
  {C.}~\bibnamefont {Mortalo}}, \bibinfo {author} {\bibfnamefont
  {E.}~\bibnamefont {Olivieri}}, \bibinfo {author} {\bibfnamefont
  {F.}~\bibnamefont {Parenti}}, \bibinfo {author} {\bibfnamefont
  {P.}~\bibnamefont {Rosa}}, \bibinfo {author} {\bibfnamefont {R.}~\bibnamefont
  {Sessoli}}, \bibinfo {author} {\bibfnamefont {L.}~\bibnamefont {Sorace}},
  \bibinfo {author} {\bibfnamefont {W.}~\bibnamefont {Wernsdorfer}}, \ and\
  \bibinfo {author} {\bibfnamefont {L.~J.}\ \bibnamefont {Zobbi}},\ }\href@noop
  {} {\bibfield  {journal} {\bibinfo  {journal} {J. Am. Chem. Soc.}\ }\textbf
  {\bibinfo {volume} {128}},\ \bibinfo {pages} {4742} (\bibinfo {year}
  {2006}{\natexlab{a}})}\BibitemShut {NoStop}%
\bibitem [{\citenamefont {Gregoli}\ \emph {et~al.}(2009)\citenamefont
  {Gregoli}, \citenamefont {Danieli}, \citenamefont {Barra}, \citenamefont
  {Neugebauer}, \citenamefont {Pellegrino}, \citenamefont {Poneti},
  \citenamefont {Sessoli},\ and\ \citenamefont {Cornia}}]{sessoli_2009}%
  \BibitemOpen
  \bibfield  {author} {\bibinfo {author} {\bibfnamefont {L.}~\bibnamefont
  {Gregoli}}, \bibinfo {author} {\bibfnamefont {C.}~\bibnamefont {Danieli}},
  \bibinfo {author} {\bibfnamefont {A.-L.}\ \bibnamefont {Barra}}, \bibinfo
  {author} {\bibfnamefont {P.}~\bibnamefont {Neugebauer}}, \bibinfo {author}
  {\bibfnamefont {G.}~\bibnamefont {Pellegrino}}, \bibinfo {author}
  {\bibfnamefont {G.}~\bibnamefont {Poneti}}, \bibinfo {author} {\bibfnamefont
  {R.}~\bibnamefont {Sessoli}}, \ and\ \bibinfo {author} {\bibfnamefont
  {A.}~\bibnamefont {Cornia}},\ }\href@noop {} {\bibfield  {journal} {\bibinfo
  {journal} {Chem.s Eur. J.}\ }\textbf {\bibinfo {volume} {15}},\ \bibinfo
  {pages} {6456} (\bibinfo {year} {2009})}\BibitemShut {NoStop}%
\bibitem [{\citenamefont {Mannini}\ \emph {et~al.}(2009)\citenamefont
  {Mannini}, \citenamefont {Pineider}, \citenamefont {Sainctavit},
  \citenamefont {Danieli}, \citenamefont {Otero}, \citenamefont
  {Sciancalepore}, \citenamefont {Talarico}, \citenamefont {Arrio},
  \citenamefont {Cornia}, \citenamefont {Gatteschi},\ and\ \citenamefont
  {Sessoli}}]{sessoli_NM_2009}%
  \BibitemOpen
  \bibfield  {author} {\bibinfo {author} {\bibfnamefont {M.}~\bibnamefont
  {Mannini}}, \bibinfo {author} {\bibfnamefont {F.}~\bibnamefont {Pineider}},
  \bibinfo {author} {\bibfnamefont {P.}~\bibnamefont {Sainctavit}}, \bibinfo
  {author} {\bibfnamefont {C.}~\bibnamefont {Danieli}}, \bibinfo {author}
  {\bibfnamefont {E.}~\bibnamefont {Otero}}, \bibinfo {author} {\bibfnamefont
  {C.}~\bibnamefont {Sciancalepore}}, \bibinfo {author} {\bibfnamefont
  {A.}~\bibnamefont {Talarico}}, \bibinfo {author} {\bibfnamefont {M.-A.}\
  \bibnamefont {Arrio}}, \bibinfo {author} {\bibfnamefont {A.}~\bibnamefont
  {Cornia}}, \bibinfo {author} {\bibfnamefont {D.}~\bibnamefont {Gatteschi}}, \
  and\ \bibinfo {author} {\bibfnamefont {R.}~\bibnamefont {Sessoli}},\
  }\href@noop {} {\bibfield  {journal} {\bibinfo  {journal} {Nat. Mater.}\
  }\textbf {\bibinfo {volume} {8}},\ \bibinfo {pages} {194} (\bibinfo {year}
  {2009})}\BibitemShut {NoStop}%
\bibitem [{\citenamefont {Mannini}\ \emph {et~al.}(2010)\citenamefont
  {Mannini}, \citenamefont {Pineider}, \citenamefont {Danieli}, \citenamefont
  {Totti}, \citenamefont {Sorace}, \citenamefont {Sainctavit}, \citenamefont
  {Arrio}, \citenamefont {Otero}, \citenamefont {Joly}, \citenamefont {Cezar},
  \citenamefont {Cornia},\ and\ \citenamefont {Sessoli}}]{sessoli10}%
  \BibitemOpen
  \bibfield  {author} {\bibinfo {author} {\bibfnamefont {M.}~\bibnamefont
  {Mannini}}, \bibinfo {author} {\bibfnamefont {F.}~\bibnamefont {Pineider}},
  \bibinfo {author} {\bibfnamefont {C.}~\bibnamefont {Danieli}}, \bibinfo
  {author} {\bibfnamefont {F.}~\bibnamefont {Totti}}, \bibinfo {author}
  {\bibfnamefont {L.}~\bibnamefont {Sorace}}, \bibinfo {author} {\bibfnamefont
  {P.}~\bibnamefont {Sainctavit}}, \bibinfo {author} {\bibfnamefont {M.~A.}\
  \bibnamefont {Arrio}}, \bibinfo {author} {\bibfnamefont {E.}~\bibnamefont
  {Otero}}, \bibinfo {author} {\bibfnamefont {L.}~\bibnamefont {Joly}},
  \bibinfo {author} {\bibfnamefont {J.~C.}\ \bibnamefont {Cezar}}, \bibinfo
  {author} {\bibfnamefont {A.}~\bibnamefont {Cornia}}, \ and\ \bibinfo {author}
  {\bibfnamefont {R.}~\bibnamefont {Sessoli}},\ }\href@noop {} {\bibfield
  {journal} {\bibinfo  {journal} {Nature}\ }\textbf {\bibinfo {volume} {468}},\
  \bibinfo {pages} {417} (\bibinfo {year} {2010})}\BibitemShut {NoStop}%
\bibitem [{\citenamefont {Zyazin}\ \emph {et~al.}(2010)\citenamefont {Zyazin},
  \citenamefont {van~den Berg}, \citenamefont {Osorio}, \citenamefont {van~der
  Zant}, \citenamefont {Konstantinidis}, \citenamefont {Leijnse}, \citenamefont
  {Wegewijs}, \citenamefont {May}, \citenamefont {Hofstetter}, \citenamefont
  {Danieli},\ and\ \citenamefont {Cornia}}]{alexander10}%
  \BibitemOpen
  \bibfield  {author} {\bibinfo {author} {\bibfnamefont {A.~S.}\ \bibnamefont
  {Zyazin}}, \bibinfo {author} {\bibfnamefont {J.~W.~G.}\ \bibnamefont {van~den
  Berg}}, \bibinfo {author} {\bibfnamefont {E.~A.}\ \bibnamefont {Osorio}},
  \bibinfo {author} {\bibfnamefont {H.~S.~J.}\ \bibnamefont {van~der Zant}},
  \bibinfo {author} {\bibfnamefont {N.~P.}\ \bibnamefont {Konstantinidis}},
  \bibinfo {author} {\bibfnamefont {M.}~\bibnamefont {Leijnse}}, \bibinfo
  {author} {\bibfnamefont {M.~R.}\ \bibnamefont {Wegewijs}}, \bibinfo {author}
  {\bibfnamefont {F.}~\bibnamefont {May}}, \bibinfo {author} {\bibfnamefont
  {W.}~\bibnamefont {Hofstetter}}, \bibinfo {author} {\bibfnamefont
  {C.}~\bibnamefont {Danieli}}, \ and\ \bibinfo {author} {\bibfnamefont
  {A.}~\bibnamefont {Cornia}},\ }\href@noop {} {\bibfield  {journal} {\bibinfo
  {journal} {Nano Letters}\ }\textbf {\bibinfo {volume} {10}},\ \bibinfo
  {pages} {3307} (\bibinfo {year} {2010})}\BibitemShut {NoStop}%
\bibitem [{\citenamefont {Zyazin}\ \emph {et~al.}(2011)\citenamefont {Zyazin},
  \citenamefont {van~der Zant}, \citenamefont {Wegewijs},\ and\ \citenamefont
  {Cornia}}]{vzant11}%
  \BibitemOpen
  \bibfield  {author} {\bibinfo {author} {\bibfnamefont {A.~S.}\ \bibnamefont
  {Zyazin}}, \bibinfo {author} {\bibfnamefont {H.~S.~J.}\ \bibnamefont {van~der
  Zant}}, \bibinfo {author} {\bibfnamefont {M.~R.}\ \bibnamefont {Wegewijs}}, \
  and\ \bibinfo {author} {\bibfnamefont {A.}~\bibnamefont {Cornia}},\
  }\href@noop {} {\bibfield  {journal} {\bibinfo  {journal} {Synth. Met.}\
  }\textbf {\bibinfo {volume} {161}},\ \bibinfo {pages} {591} (\bibinfo {year}
  {2011})}\BibitemShut {NoStop}%
\bibitem [{\citenamefont {Burzur\'\i}\ \emph {et~al.}(2012)\citenamefont
  {Burzur\'\i}, \citenamefont {Zyazin}, \citenamefont {Cornia},\ and\
  \citenamefont {van~der Zant}}]{vzant12}%
  \BibitemOpen
  \bibfield  {author} {\bibinfo {author} {\bibfnamefont {E.}~\bibnamefont
  {Burzur\'\i}}, \bibinfo {author} {\bibfnamefont {A.~S.}\ \bibnamefont
  {Zyazin}}, \bibinfo {author} {\bibfnamefont {A.}~\bibnamefont {Cornia}}, \
  and\ \bibinfo {author} {\bibfnamefont {H.~S.~J.}\ \bibnamefont {van~der
  Zant}},\ }\href {\doibase 10.1103/PhysRevLett.109.147203} {\bibfield
  {journal} {\bibinfo  {journal} {Phys. Rev. Lett.}\ }\textbf {\bibinfo
  {volume} {109}},\ \bibinfo {pages} {147203} (\bibinfo {year}
  {2012})}\BibitemShut {NoStop}%
\bibitem [{\citenamefont {Michalak}\ \emph {et~al.}(2010)\citenamefont
  {Michalak}, \citenamefont {Canali}, \citenamefont {Pederson}, \citenamefont
  {Paulsson},\ and\ \citenamefont {Benza}}]{carlo10}%
  \BibitemOpen
  \bibfield  {author} {\bibinfo {author} {\bibfnamefont {{\L}.}~\bibnamefont
  {Michalak}}, \bibinfo {author} {\bibfnamefont {C.~M.}\ \bibnamefont
  {Canali}}, \bibinfo {author} {\bibfnamefont {M.~R.}\ \bibnamefont
  {Pederson}}, \bibinfo {author} {\bibfnamefont {M.}~\bibnamefont {Paulsson}},
  \ and\ \bibinfo {author} {\bibfnamefont {V.~G.}\ \bibnamefont {Benza}},\
  }\href {\doibase 10.1103/PhysRevLett.104.017202} {\bibfield  {journal}
  {\bibinfo  {journal} {Phys. Rev. Lett.}\ }\textbf {\bibinfo {volume} {104}},\
  \bibinfo {pages} {017202} (\bibinfo {year} {2010})}\BibitemShut {NoStop}%
\bibitem [{\citenamefont {Barraza-Lopez}\ \emph {et~al.}(2009)\citenamefont
  {Barraza-Lopez}, \citenamefont {Park}, \citenamefont {Garc{\'i}a-Su\'arez},\
  and\ \citenamefont {Ferrer}}]{ferrer_prl2009}%
  \BibitemOpen
  \bibfield  {author} {\bibinfo {author} {\bibfnamefont {S.}~\bibnamefont
  {Barraza-Lopez}}, \bibinfo {author} {\bibfnamefont {K.}~\bibnamefont {Park}},
  \bibinfo {author} {\bibfnamefont {V.}~\bibnamefont {Garc{\'i}a-Su\'arez}}, \
  and\ \bibinfo {author} {\bibfnamefont {J.}~\bibnamefont {Ferrer}},\ }\href
  {\doibase 10.1103/PhysRevLett.102.246801} {\bibfield  {journal} {\bibinfo
  {journal} {Phys. Rev. Lett.}\ }\textbf {\bibinfo {volume} {102}},\ \bibinfo
  {pages} {246801} (\bibinfo {year} {2009})}\BibitemShut {NoStop}%
\bibitem [{\citenamefont {Rostamzadeh~Renani}\ and\ \citenamefont
  {Kirczenow}(2012)}]{renani2012}%
  \BibitemOpen
  \bibfield  {author} {\bibinfo {author} {\bibfnamefont {F.}~\bibnamefont
  {Rostamzadeh~Renani}}\ and\ \bibinfo {author} {\bibfnamefont
  {G.}~\bibnamefont {Kirczenow}},\ }\href {\doibase 10.1103/PhysRevB.85.245415}
  {\bibfield  {journal} {\bibinfo  {journal} {Phys. Rev. B}\ }\textbf {\bibinfo
  {volume} {85}},\ \bibinfo {pages} {245415} (\bibinfo {year}
  {2012})}\BibitemShut {NoStop}%
\bibitem [{\citenamefont {Pederson}\ and\ \citenamefont
  {Khanna}(1999)}]{mark99}%
  \BibitemOpen
  \bibfield  {author} {\bibinfo {author} {\bibfnamefont {M.~R.}\ \bibnamefont
  {Pederson}}\ and\ \bibinfo {author} {\bibfnamefont {S.~N.}\ \bibnamefont
  {Khanna}},\ }\href {\doibase 10.1103/PhysRevB.60.9566} {\bibfield  {journal}
  {\bibinfo  {journal} {Phys. Rev. B}\ }\textbf {\bibinfo {volume} {60}},\
  \bibinfo {pages} {9566} (\bibinfo {year} {1999})}\BibitemShut {NoStop}%
\bibitem [{\citenamefont {Kortus}\ \emph {et~al.}(2003)\citenamefont {Kortus},
  \citenamefont {Pederson}, \citenamefont {Baruah}, \citenamefont {Bernstein},\
  and\ \citenamefont {Hellberg}}]{mark03}%
  \BibitemOpen
  \bibfield  {author} {\bibinfo {author} {\bibfnamefont {J.}~\bibnamefont
  {Kortus}}, \bibinfo {author} {\bibfnamefont {M.~R.}\ \bibnamefont
  {Pederson}}, \bibinfo {author} {\bibfnamefont {T.}~\bibnamefont {Baruah}},
  \bibinfo {author} {\bibfnamefont {N.}~\bibnamefont {Bernstein}}, \ and\
  \bibinfo {author} {\bibfnamefont {C.}~\bibnamefont {Hellberg}},\ }\href@noop
  {} {\bibfield  {journal} {\bibinfo  {journal} {Polyhedron}\ }\textbf
  {\bibinfo {volume} {22}},\ \bibinfo {pages} {1871} (\bibinfo {year}
  {2003})}\BibitemShut {NoStop}%
\bibitem [{\citenamefont {Postnikov}\ \emph {et~al.}(2006)\citenamefont
  {Postnikov}, \citenamefont {Kortus},\ and\ \citenamefont
  {Pederson}}]{mark06}%
  \BibitemOpen
  \bibfield  {author} {\bibinfo {author} {\bibfnamefont {A.~V.}\ \bibnamefont
  {Postnikov}}, \bibinfo {author} {\bibfnamefont {J.}~\bibnamefont {Kortus}}, \
  and\ \bibinfo {author} {\bibfnamefont {M.~R.}\ \bibnamefont {Pederson}},\
  }\href@noop {} {\bibfield  {journal} {\bibinfo  {journal} {Phys. Stat. Sol.
  (b)}\ }\textbf {\bibinfo {volume} {243}},\ \bibinfo {pages} {2533} (\bibinfo
  {year} {2006})}\BibitemShut {NoStop}%
\bibitem [{\citenamefont {Pederson}\ and\ \citenamefont {Baruah}(2007)}]{hmm}%
  \BibitemOpen
  \bibfield  {author} {\bibinfo {author} {\bibfnamefont {M.~R.}\ \bibnamefont
  {Pederson}}\ and\ \bibinfo {author} {\bibfnamefont {T.}~\bibnamefont
  {Baruah}},\ }\href@noop {} {\emph {\bibinfo {title} {Handbook of Magnetism
  and Magnetic Materials}}},\ edited by\ \bibinfo {editor} {\bibfnamefont
  {S.}~\bibnamefont {Parkin}}\ and\ \bibinfo {editor} {\bibfnamefont
  {H.}~\bibnamefont {Kronmullerr}}\ (\bibinfo  {publisher} {J. Wiley and Sons,
  London},\ \bibinfo {year} {2007})\ Chap.~\bibinfo {chapter} {9}\BibitemShut
  {NoStop}%
\bibitem [{\citenamefont {Baruah}\ and\ \citenamefont
  {Pederson}(2003)}]{baruah}%
  \BibitemOpen
  \bibfield  {author} {\bibinfo {author} {\bibfnamefont {T.}~\bibnamefont
  {Baruah}}\ and\ \bibinfo {author} {\bibfnamefont {M.~R.}\ \bibnamefont
  {Pederson}},\ }\href@noop {} {\bibfield  {journal} {\bibinfo  {journal} {Int.
  J. of Quant. Chem.}\ }\textbf {\bibinfo {volume} {93}},\ \bibinfo {pages}
  {324} (\bibinfo {year} {2003})}\BibitemShut {NoStop}%
\bibitem [{\citenamefont {Park}\ and\ \citenamefont {Pederson}(2004)}]{park}%
  \BibitemOpen
  \bibfield  {author} {\bibinfo {author} {\bibfnamefont {K.}~\bibnamefont
  {Park}}\ and\ \bibinfo {author} {\bibfnamefont {M.~R.}\ \bibnamefont
  {Pederson}},\ }\href@noop {} {\bibfield  {journal} {\bibinfo  {journal}
  {Phys. Rev. B}\ }\textbf {\bibinfo {volume} {70}},\ \bibinfo {pages} {054414}
  (\bibinfo {year} {2004})}\BibitemShut {NoStop}%
\bibitem [{\citenamefont {van Wüllen}(2009)}]{vanwullen1}%
  \BibitemOpen
  \bibfield  {author} {\bibinfo {author} {\bibfnamefont {C.}~\bibnamefont {van
  Wüllen}},\ }\href@noop {} {\bibfield  {journal} {\bibinfo  {journal} {J.
  Chem. Phys.}\ }\textbf {\bibinfo {volume} {130}},\ \bibinfo {pages} {194109}
  (\bibinfo {year} {2009})}\BibitemShut {NoStop}%
\bibitem [{\citenamefont {S.~Schmitt}\ and\ \citenamefont
  {v~Wullen}(2011)}]{vanwullen2}%
  \BibitemOpen
  \bibfield  {author} {\bibinfo {author} {\bibfnamefont {P.~J.}\ \bibnamefont
  {S.~Schmitt}}\ and\ \bibinfo {author} {\bibfnamefont {C.}~\bibnamefont
  {v~Wullen}},\ }\href@noop {} {\bibfield  {journal} {\bibinfo  {journal} {J.
  Chem. Phys.}\ }\textbf {\bibinfo {volume} {134}},\ \bibinfo {pages} {194113}
  (\bibinfo {year} {2011})}\BibitemShut {NoStop}%
\bibitem [{\citenamefont {Janak}(1978)}]{janak}%
  \BibitemOpen
  \bibfield  {author} {\bibinfo {author} {\bibfnamefont {J.~F.}\ \bibnamefont
  {Janak}},\ }\href@noop {} {\bibfield  {journal} {\bibinfo  {journal} {Phys.
  Rev. B}\ }\textbf {\bibinfo {volume} {18}},\ \bibinfo {pages} {7165}
  (\bibinfo {year} {1978})}\BibitemShut {NoStop}%
\bibitem [{\citenamefont {Pederson}\ and\ \citenamefont
  {Jackson}(1991)}]{pedjak91}%
  \BibitemOpen
  \bibfield  {author} {\bibinfo {author} {\bibfnamefont {M.~R.}\ \bibnamefont
  {Pederson}}\ and\ \bibinfo {author} {\bibfnamefont {K.~A.}\ \bibnamefont
  {Jackson}},\ }\href@noop {} {\bibfield  {journal} {\bibinfo  {journal} {Phys.
  Rev. B}\ }\textbf {\bibinfo {volume} {43}},\ \bibinfo {pages} {7312}
  (\bibinfo {year} {1991})}\BibitemShut {NoStop}%
\bibitem [{\citenamefont {Pederson}\ and\ \citenamefont
  {Jackson}(1990)}]{mark90_1}%
  \BibitemOpen
  \bibfield  {author} {\bibinfo {author} {\bibfnamefont {M.~R.}\ \bibnamefont
  {Pederson}}\ and\ \bibinfo {author} {\bibfnamefont {K.~A.}\ \bibnamefont
  {Jackson}},\ }\href@noop {} {\bibfield  {journal} {\bibinfo  {journal} {Phys.
  Rev. B}\ }\textbf {\bibinfo {volume} {41}},\ \bibinfo {pages} {7453}
  (\bibinfo {year} {1990})}\BibitemShut {NoStop}%
\bibitem [{\citenamefont {Jackson}\ and\ \citenamefont
  {Pederson}(1990)}]{mark90_2}%
  \BibitemOpen
  \bibfield  {author} {\bibinfo {author} {\bibfnamefont {K.~A.}\ \bibnamefont
  {Jackson}}\ and\ \bibinfo {author} {\bibfnamefont {M.~R.}\ \bibnamefont
  {Pederson}},\ }\href@noop {} {\bibfield  {journal} {\bibinfo  {journal}
  {Phys. Rev. B}\ }\textbf {\bibinfo {volume} {42}},\ \bibinfo {pages} {3276}
  (\bibinfo {year} {1990})}\BibitemShut {NoStop}%
\bibitem [{\citenamefont {Perdew}\ \emph {et~al.}(1996)\citenamefont {Perdew},
  \citenamefont {Burke},\ and\ \citenamefont {Ernzerhof}}]{perdew}%
  \BibitemOpen
  \bibfield  {author} {\bibinfo {author} {\bibfnamefont {J.~P.}\ \bibnamefont
  {Perdew}}, \bibinfo {author} {\bibfnamefont {K.}~\bibnamefont {Burke}}, \
  and\ \bibinfo {author} {\bibfnamefont {M.}~\bibnamefont {Ernzerhof}},\
  }\href@noop {} {\bibfield  {journal} {\bibinfo  {journal} {Phys. Rev. Lett.}\
  }\textbf {\bibinfo {volume} {77}},\ \bibinfo {pages} {3865} (\bibinfo {year}
  {1996})}\BibitemShut {NoStop}%
\bibitem [{\citenamefont {Accorsi}\ \emph
  {et~al.}(2006{\natexlab{b}})\citenamefont {Accorsi}, \citenamefont {Barra},
  \citenamefont {Caneschi}, \citenamefont {Chastanet}, \citenamefont {Cornia},
  \citenamefont {Fabretti}, \citenamefont {Gatteschi}, \citenamefont
  {Mortalò}, \citenamefont {Olivieri}, \citenamefont {Parenti}, \citenamefont
  {Rosa}, \citenamefont {Sessoli}, \citenamefont {Sorace}, \citenamefont
  {Wernsdorfer},\ and\ \citenamefont {Zobbi}}]{accors06}%
  \BibitemOpen
  \bibfield  {author} {\bibinfo {author} {\bibfnamefont {S.}~\bibnamefont
  {Accorsi}}, \bibinfo {author} {\bibfnamefont {A.-L.}\ \bibnamefont {Barra}},
  \bibinfo {author} {\bibfnamefont {A.}~\bibnamefont {Caneschi}}, \bibinfo
  {author} {\bibfnamefont {G.}~\bibnamefont {Chastanet}}, \bibinfo {author}
  {\bibfnamefont {A.}~\bibnamefont {Cornia}}, \bibinfo {author} {\bibfnamefont
  {A.~C.}\ \bibnamefont {Fabretti}}, \bibinfo {author} {\bibfnamefont
  {D.}~\bibnamefont {Gatteschi}}, \bibinfo {author} {\bibfnamefont
  {C.}~\bibnamefont {Mortalò}}, \bibinfo {author} {\bibfnamefont
  {E.}~\bibnamefont {Olivieri}}, \bibinfo {author} {\bibfnamefont
  {F.}~\bibnamefont {Parenti}}, \bibinfo {author} {\bibfnamefont
  {P.}~\bibnamefont {Rosa}}, \bibinfo {author} {\bibfnamefont {R.}~\bibnamefont
  {Sessoli}}, \bibinfo {author} {\bibfnamefont {L.}~\bibnamefont {Sorace}},
  \bibinfo {author} {\bibfnamefont {W.}~\bibnamefont {Wernsdorfer}}, \ and\
  \bibinfo {author} {\bibfnamefont {L.}~\bibnamefont {Zobbi}},\ }\href@noop {}
  {\bibfield  {journal} {\bibinfo  {journal} {J. Am. Chem. Soc.}\ }\textbf
  {\bibinfo {volume} {128}},\ \bibinfo {pages} {4742} (\bibinfo {year}
  {2006}{\natexlab{b}})}\BibitemShut {NoStop}%
\bibitem [{\citenamefont {Barra}\ \emph {et~al.}(1999)\citenamefont {Barra},
  \citenamefont {Caneschi}, \citenamefont {Cornia}, \citenamefont {Fabrizi~de
  Biani}, \citenamefont {Gatteschi}, \citenamefont {Sangregorio}, \citenamefont
  {Sessoli},\ and\ \citenamefont {Sorace}}]{barra99}%
  \BibitemOpen
  \bibfield  {author} {\bibinfo {author} {\bibfnamefont {A.~L.}\ \bibnamefont
  {Barra}}, \bibinfo {author} {\bibfnamefont {A.}~\bibnamefont {Caneschi}},
  \bibinfo {author} {\bibfnamefont {A.}~\bibnamefont {Cornia}}, \bibinfo
  {author} {\bibfnamefont {F.}~\bibnamefont {Fabrizi~de Biani}}, \bibinfo
  {author} {\bibfnamefont {D.}~\bibnamefont {Gatteschi}}, \bibinfo {author}
  {\bibfnamefont {C.}~\bibnamefont {Sangregorio}}, \bibinfo {author}
  {\bibfnamefont {R.}~\bibnamefont {Sessoli}}, \ and\ \bibinfo {author}
  {\bibfnamefont {L.}~\bibnamefont {Sorace}},\ }\href@noop {} {\bibfield
  {journal} {\bibinfo  {journal} {J. Am. Chem. Soc.}\ }\textbf {\bibinfo
  {volume} {121}},\ \bibinfo {pages} {5302} (\bibinfo {year}
  {1999})}\BibitemShut {NoStop}%
\bibitem [{\citenamefont {Ribas-Arino}\ \emph {et~al.}(2005)\citenamefont
  {Ribas-Arino}, \citenamefont {Baruah},\ and\ \citenamefont
  {Pederson}}]{mark05}%
  \BibitemOpen
  \bibfield  {author} {\bibinfo {author} {\bibfnamefont {J.}~\bibnamefont
  {Ribas-Arino}}, \bibinfo {author} {\bibfnamefont {T.}~\bibnamefont {Baruah}},
  \ and\ \bibinfo {author} {\bibfnamefont {M.~R.}\ \bibnamefont {Pederson}},\
  }\href {\doibase 10.1063/1.1961367} {\bibfield  {journal} {\bibinfo
  {journal} {J. Chem. Phys.}\ }\textbf {\bibinfo {volume} {123}},\ \bibinfo
  {eid} {044303} (\bibinfo {year} {2005})}\BibitemShut {NoStop}%
\bibitem [{\citenamefont {Li}\ \emph {et~al.}(2003)\citenamefont {Li},
  \citenamefont {Li}, \citenamefont {Zhai},\ and\ \citenamefont {Wang}}]{li03}%
  \BibitemOpen
  \bibfield  {author} {\bibinfo {author} {\bibfnamefont {J.}~\bibnamefont
  {Li}}, \bibinfo {author} {\bibfnamefont {X.}~\bibnamefont {Li}}, \bibinfo
  {author} {\bibfnamefont {H.-J.}\ \bibnamefont {Zhai}}, \ and\ \bibinfo
  {author} {\bibfnamefont {L.-S.}\ \bibnamefont {Wang}},\ }\href@noop {}
  {\bibfield  {journal} {\bibinfo  {journal} {Science}\ }\textbf {\bibinfo
  {volume} {299}},\ \bibinfo {pages} {864} (\bibinfo {year}
  {2003})}\BibitemShut {NoStop}%
\bibitem [{Note1()}]{Note1}%
  \BibitemOpen
  \bibinfo {note} {The fact that in Table \ref {ch:lead} the spin magnetic
  moment of the $Q= -1$ charge state is not exactly equal to an integer is due
  to the fractional occupancy (approximately 15\%) of the spin-down LUMO in our
  calculations, carried out with a finite smearing of the Fermi-Dirac
  distribution.}\BibitemShut {Stop}%
\bibitem [{Note2()}]{Note2}%
  \BibitemOpen
  \bibinfo {note} {Note however that the amplitude of the LUMO wavefunction is
  quite small in the central region. Therefore, an electron tunneling in from
  one of the leads would still find a bottleneck when tunneling out to the
  other lead.}\BibitemShut {Stop}%
\bibitem [{Note3()}]{Note3}%
  \BibitemOpen
  \bibinfo {note} {Note that in Ref.~\protect \rev@citealpnum {alexander10},
  the results reported for the anion and cation were obtained for sample A and
  B respectively.}\BibitemShut {Stop}%
\bibitem [{Note4()}]{Note4}%
  \BibitemOpen
  \bibinfo {note} {In general, in Ref.~\protect \rev@citealpnum {vzant12} it is
  found that upon reduction, either from $N \rightarrow N+1$ or from $N-1
  \rightarrow N$, the spin $S$ always {\protect \it decreases} and the
  anisotropy parameter $D$, defined via $\Delta = D S^2$, always {\protect \it
  increases}. The results of our calculations (see Table~\ref {ch:lead}) show
  that both the anion and the cation have preferably $S= 11/2$ for a lead of
  Type-2, whereas $S= 9/2$ for a lead of Type-1. However, states with swapped
  spin configurations $S= 11/2 \leftrightarrow S= 9/2$ are quite close in
  energy for both charged states.}\BibitemShut {Stop}%
\end{thebibliography}%

\end{document}